\documentclass[12pt]{article}
\usepackage{epsf}
\usepackage{graphicx}
\usepackage{amsfonts}
\def\hybrid{\topmargin 0pt      \oddsidemargin 0pt
	\headheight 0pt \headsep 0pt
	\textheight 9in         
	\textwidth 6.25in       
	\marginparwidth .875in
	\parskip 5pt plus 1pt   \jot = 1.5ex}

\catcode`\@=11
\def\marginnote#1{}
\newcount\hour
\newcount\minute
\newtoks\amorpm
\hour=\time\divide\hour by60
\minute=\time{\multiply\hour by60 \global\advance\minute by-\hour}
\edef\standardtime{{\ifnum\hour<12 \global\amorpm={am}%
	\else\global\amorpm={pm}\advance\hour by-12 \fi
	\ifnum\hour=0 \hour=12 \fi
	\number\hour:\ifnum\minute<10 0\fi\number\minute\the\amorpm}}
\edef\militarytime{\number\hour:\ifnum\minute<10 0\fi\number\minute}

\def\draftlabel#1{{\@bsphack\if@filesw {\let\thepage\relax
   \xdef\@gtempa{\write\@auxout{\string
      \newlabel{#1}{{\@currentlabel}{\thepage}}}}}\@gtempa
   \if@nobreak \ifvmode\nobreak\fi\fi\fi\@esphack}
	\gdef\@eqnlabel{#1}}
\def\@eqnlabel{}
\def\@vacuum{}
\def\draftmarginnote#1{\marginpar{\raggedright\scriptsize\tt#1}}

\def\draft{\oddsidemargin -.5truein
	\def\@oddfoot{\sl preliminary draft \hfil
	\rm\thepage\hfil\sl\today\quad\militarytime}
	\let\@evenfoot\@oddfoot \overfullrule 3pt
	\let\label=\draftlabel
\let\marginnote=\draftmarginnote
	\let\marginnote=\draftmarginnote
   \def\@eqnnum{(\theequation)\rlap{\kern\marginparsep\tt\@eqnlabel}%
\global\let\@eqnlabel\@vacuum}  }

\def\numberbysection{\@addtoreset{equation}{section}
	\def\theequation{\thesection.\arabic{equation}}}

\def\underline#1{\relax\ifmmode\@@underline#1\else
	$\@@underline{\hbox{#1}}$\relax\fi}

\def\titlepage{\@restonecolfalse\if@twocolumn\@restonecoltrue\onecolumn
     \else \newpage \fi \thispagestyle{empty}\c@page\z@
	\def\thefootnote{\fnsymbol{footnote}} }

\def\endtitlepage{\if@restonecol\twocolumn \else  \fi
	\def\thefootnote{\arabic{footnote}}
	\setcounter{footnote}{0}}  
\catcode`@=12
\relax

\def\beq{\begin{equation}}
\def\eeq{\end{equation}}
\def\bea{\begin{eqnarray}}
\def\eea{\end{eqnarray}}

\relax
\hybrid

\newcommand{\wb}[2]{W\!B_{#1}^{#2}}
\newcommand{\z}[2]{Z^{(2)}_{#1}(#2)}
\newcommand{\ud}{\mathrm{d}}

\makeatletter
\renewcommand\theequation{\thesection.\arabic{equation}}
\@addtoreset{equation}{section}
\makeatother

\begin{document}

\begin{titlepage}
\begin{center}
February~2007 \hfill . \\[.5in]
{\large\bf Renormalization group flows for the second $Z_{N}$ parafermionic field theory for N odd.} 
\\[.5in] 
{\bf Vladimir S.~Dotsenko 
 and Benoit Estienne}\\[.2in]
 {\it Laboratoire de Physique Th\'eorique 
et Hautes Energies}\footnote{Unit\'e Mixte de Recherche UMR 7589}\\

 {\it Universit{\'e} Pierre et Marie Curie, Paris-6; CNRS\\
              Universit{\'e} Denis Diderot, Paris-7\\
               Bo\^{\i}te 126, Tour 25, 5\`eme {\'e}tage,
               4 place Jussieu, F-75252 Paris Cedex 05, France.}\\
 
{\it dotsenko@lpthe.jussieu.fr,     estienne@lpthe.jussieu.fr}\\

\end{center}

\underline{Abstract.}

Using the renormalization group approach, the Coulomb gas and the coset techniques, 
the effect of slightly relevant perturbations is studied 
for the second parafermionic field theory with the symmetry $Z_{N}$, for N odd. 
New fixed points are found and classified.

\end{titlepage}

\newpage

Parafermionic conformal field theories describe systems enjoying conformal symmetry and a cyclic symmetry $Z_{N}$.

The first series of parafermionic conformal field theories appeared in 1985 \cite{ref1}. Since then they have been well studied and applied in various domains \cite{ref2,ref3,ref4}. The second parafermionic series $Z_{N}^{(2)}$ has been developed fairly recently [5-8] and it still awaits its applications.

In the case of the first series $Z_{N}^{(1)}$, to a given $N$ is associated a single conformal theory. This is different for the second series: for a given $N$, there exist an infinity of unitary conformal theories $Z_{N}^{(2)}(p)$, with $p=N-1,N-2,...$. These theories correspond to degenerate representations of the corresponding parafermionic chiral algebra. They are much more rich in their content of physical fields, as compared to the theories of the first series. They are also much more complicated. But, on the other hand, the presence of the parameter $p$, for a given $Z_{N}$, opens a way to reliable perturbative studies. It allows in particular to study the renormalization group flows in the space of these conformal theory models, under various perturbations.

In the theory of second order phase transition, it is widely accepted that fixed points of the renormalization group should be described by conformal field theories. Saying it differently, a CFT describes the critical point of some statistical system. In order to investigate the behavior of the renormalization group in the vicinity of a fixed point, it is useful to study the effects of slightly relevant perturbations of the correponding conformal field theory.

 In this paper we shall present results for the renormalization group flows of the $\z{N}{p}$ theories with $N$ odd, being perturbed by two slightly relevant fields. In a previous letter \cite{refletter} we have studied the case of the $\z{5}{p}$ theories. These results are here generalized to any odd integer $N$, and more details are given.

The details of the $Z_{N}^{(2)}$ parafermionic theory with $N$ odd could be found in \cite{ref6}. The $q$ charge of $Z_{N}$ takes values in 
$Z_{N}$, so that in the Kac table of this theory one finds the $Z_{N}^{(2)}$ neutral fields, of $q=0$, the $q = 1 \dots N-1 $ doublets, and the $Z_{2}$ disorder fields. The symmetry of the theory is actually $D_{N}$, which is made of $Z_{N}$ rotations and the $Z_{2}$ reflections in N different axes. These last symmetry elements amount to the charge conjugation symmetry: $q\rightarrow -q$.

In this paper we will treat the case $N$ odd. The first non trivial $Z_{N}^{(2)}$ theory with $N$ odd is $Z^{(2)}_3$, and its renormalization by slightly relevant fields has already been treated in \cite{ref11,ref12}. So we will analyse the case $N\geq 5$. We expect different results from the case $N=3$ since \beq \z{3}{p} = \frac{SU_k(2) \otimes SU_4(2)}{SU_{k+4}(2)} \eeq is a $SU(2)$ coset with a shift parameter 4, while the $N \geq 5$ theories are $SO(N)$ cosets with a shift parameter 2 : \beq \z{N}{p} = \frac{SO_k(N) \otimes SO_2(N)}{SO_{k+2}(N)} \eeq





\section{Perturbing $Z_{N}^{(2)}$}



A conformal theory can be seen as the field theory describing the critical point of some statistical system, i.e. the fixed point of the renomalization group. In order to get some insights into the neighborhood of this critical point, one can study the effects of slightly relevant perturbations of the correponding conformal field theory. To do so, one need to identify a set of slightly pertinent fields : spinless fields with anomalous dimension $d = 2 - 2 \eta $, $\eta \ll 1$. The action takes the following form : 

\beq
{\bf A}={\bf A}_{0}+\sum_i g_i\int \ud^{2}x \ \Phi_i(x) 
\eeq

For the perturbed theory to be renormalizable these fields should not produce additional slightly relevant fields when fused together : the set of slighlty relevant fields must close with respect to fusion rules.

Considering slightly relevant fields allows to use perturbation theory. In the leading approximation in $\eta$, the renormalization group equations $\delta g_i = \beta_i(g) \delta \xi $ are obtained directly from the relevant fields fusion rules \cite{ref9}.

We want to perturb the second parafermionic theory, which we will denote as $Z_{N}^{(2)}$. First we need to identify a set of slightly relevant fields of this theory, which close by the operator algebra. Slightly relevant fields are fields with conformal dimension $\Delta = \overline{\Delta} = 1 - \eta $, with $\eta \ll 1$ being a small positive parameter. Since there are no fields with negative dimension, slightly relevant fields are necessarily Virasoro primary. Note that does not necessarily mean $Z_N$ primary. 

Perturbatively well controled domain of $Z_{N}^{(2)}(p)$ theories is that of $ p\gg 1$, giving a small parameter $\epsilon\sim 1/p$. This is similar to the original perturbative renormalization group treatment of minimal models for Virasoro algebra based conformal theory \cite{ref9,ref10}.

Since we want to conserve the $Z_N$ symmetry, we demand these fields to be neutral w.r.t. $\mathbb{Z}_N$. Slightly relevant neutral fields can be of 2 sorts : 

\begin{itemize}

\item a $Z_N$ primary, singlet (q=0). We will denote these fields as $S$,

\item a $Z_N$ neutral descendant of a doublet : $A = \Psi^{q_1}_{-x_1} \ldots \Psi^{q_n}_{-x_n}D^{q}$ , with the neutrality condition $\sum_i q_i + q = 0 $ mod $N$ 

\end{itemize}

\subsection{Singlet S}

The Kac formula for $\z{N}{p}$ has been given in \cite{ref6} ; it can be found in Appendix \ref{appendixZN}. The conformal dimension of a primary singlet $S_{(\vec{n} \mid \vec{n}')}$ is given by : 

\bea
\Delta^{S}_{(\vec{n} \mid \vec{n}')} & = & \frac{((p+2)\vec{n} - p\vec{n}')^2 - 4\vec{\rho}^2}{4p(p+2)} \cr
& = & \frac{(\vec{n}-\vec{n}')^2}{4} - \frac{\vec{n}'^{2}-\vec{n}^2}{2} \epsilon + \mathcal{O}(\epsilon^2) 
\eea

There are infinitely many solutions to $\Delta = 1 - \eta$ as $ p \rightarrow \infty $. But we want a closed algebra of slightly relevant fields, with a bounded number of fields that does not depend on $p$. It is similar to the case of minimal models in which there are many slightly relevant field : all the $\phi_{n,n+3}$. But the field $\phi_{1,3}$ alone forms a closed algebra : its fusion does not generate the other fields $\phi_{n,n+3}$ with $n>1$ simply because the $\alpha_{+}$ side of $\phi_{1,3}$ is trivial (n=1). We will do a similar treatment here. In order to help ensuring the closing of the fields, we will impose the following condition : $\vec{n} = (1,1,\ldots,1)$, i.e. we demand the $\alpha_+ $ side to be trivial.

There remains one unique singlet : 
\bea
S & = & S_{(1,1,\ldots 1\mid 3,1,1,\ldots 1)} \\
\Delta_S & = & 1 - N\epsilon + \mathcal{O}(\epsilon^2)
\eea

\subsection{Fundamental descendant of a doublet $D^{Q}$}

By fundamental descendant we mean a $Z_N$ descendant that is still Virasoro primary. The doublets $\mathcal{D}^{Q=2q}$, $Q = 0,1,\ldots \frac{N-1}{2}$ have a non trivial boundary term in their dimension.
Any $Z_N$ fundamental descendant $A = \Psi^{q_1}_{-x_1} \ldots \Psi^{q_n}_{-x_n}D^{Q}$ that satisfies the neutrality condition $\sum_i 2q_i + Q = 0 $ mod $N$ necessarily has a gap $ \sum_i x_i $ equal to the fundamental gap $ \delta_Q = \frac{Q(Q-N)}{2N}$ mod[1]. The conformal dimension of such a descendant is (cf appendix \ref{appendix_pertinentfields}): 

\beq
\Delta^{A}_{(1, 1,\ldots \mid \vec{n}')}  =  \frac{((p+2)\vec{\rho} - p\vec{n}')^2 - 4\vec{\rho}^2}{4p(p+2)} + B_Q + \delta_Q
\eeq

We want $\Delta^{A}$ to be smaller than 1. This condition drastically reduces the admissible fields. The details of the analysis are given in apprendix \ref{appendix_pertinentfields}, and we find that there is one single neutral descendant of a doublet that is slightly relevant :

\beq
\begin{array}{ll}
 A  & = \left\{	\begin{array}{ll}
A^{-1}_{-\frac{2}{5}}\Phi_{(1 1  \mid 1 3 )} & \textrm{for }N=5\\
A^{-1}_{-\frac{2}{N}}\Phi_{(1 1 1 \ldots \mid 1 2 1 \ldots)} & \textrm{for } N \geq 7
	\end{array} \right.
\end{array}
\eeq






Finally we have two $Z_{N}$ neutral fields which are slightly relevant. Since these are the only ones with a trivial $\alpha_+$ side, they necessarily form a closed algebra amongst all the slightly relevant fields. They are : 

\beq
\begin{array}{ll}
 S  & =  \Phi_{(1 1 1 \ldots \mid 3 1 1 \ldots)} \\
A  & = \left\{	\begin{array}{ll}
A^{-1}_{-\frac{2}{5}}\Phi_{(1 1  \mid 1 3 )} & \textrm{for }N=5\\
A^{-1}_{-\frac{2}{N}}\Phi_{(1 1 1 \ldots \mid 1 2 1 \ldots)} & \textrm{for } N \geq 7
	\end{array} \right.
\end{array}
\label{relevantfields}
\eeq

We observe that the case $N = 5$ is slightly different from the case $N \geq 7$. This is somewhat conventionnal, caused by the notations adopted in the labeling of the primary fields (we could have redefined $\tilde{\omega}_{r} = 2 \omega_r$ to absorb it). But we chose to keep the usual notations for the $B_r$ weights. In the following we will treat preferably the case $N \geq 7$ when we explicitly write the field $A$, the case $N=5$ being treated exactly the same way. In particular the final results hold in both cases : we observe the same phase diagram, with the same structure for the fixed points.

The conformal dimensions of the fields (\ref{relevantfields}) are given by (\ref{KacZN}). The dimensions of the fields $S$ and $A$ have the following values:

\beq
\begin{array}{lll}
\Delta_{S} & = & 1-N\epsilon \\
\Delta_{A} & = & 1-(N-2)\epsilon
\end{array} 
\eeq

We have defind $\epsilon$ as follows:
\beq
\epsilon=\frac{1}{p+2}\simeq\frac{1}{p}
\eeq

Perturbing with the fields $S$ and $A$ corresponds to taking the action of the theory in the form:
\beq
{\bf A}={\bf A}_{0}+\frac{2g}{\pi}\int d^{2}xS(x)+\frac{2h}{\pi}\int d^{2}x A(x) 
\eeq
where $g$ and $h$ are the corresponding coupling constants; 
the additional factors  $\frac{2}{\pi}$  are added to simplify the coefficients 
of the renormalization group equations which follow;  ${\bf A}_{0}$ is assumed 
to be the action of the unperturbed $\z{N}{p}$ conformal theory.

 It will be shown below that the operator algebra of the fields $S$ and $A$ is of the form:
\beq
S(x')S(x)=\frac{\mathcal{D}_{SSA}}{|x'-x|^{4\Delta_{S}-2\Delta_{A}}}A(x)+...
\label{eqn9}
\eeq
\beq
A(x')A(x)=\frac{\mathcal{D}_{AAA}}{|x'-x|^{2\Delta_{A}}} A(x)+...
\label{eqn10}
\eeq
\beq
S(x')A(x)=\frac{\mathcal{D}_{SSA}}{|x'-x|^{2\Delta_{A}}} S(x)+...
\label{eqn11}
\eeq
Only the fields which are relevant for the renormalization group flows are shown explicitly in the r.h.s. of the equations (\ref{eqn9})-(\ref{eqn11}). For instance, the identity operator is not shown in the r.h.s. of (\ref{eqn9}) and (\ref{eqn10}) while it is naturally present there.
The operator algebra constants in (\ref{eqn9}) and (\ref{eqn11}) should 
obviously be equal, as the two equations could be related to 
a single correlation function     $<S(x_{1})S(x_{2})A(x_{3})>$.

Assuming the operator algebra expansions in (9)-(11), one finds, in a standard way, the following renormalization group equations for the couplings $g$ and $h$:
\bea
\frac{dg}{d\xi}& = & 2 N \epsilon \  g-4\mathcal{D}_{SSA}\ g\ h \\
\frac{dh}{d\xi}& = & 2 (N-2) \epsilon \  h-2\mathcal{D}_{AAA}\ h^{2}-2\mathcal{D}_{SSA}\ g^{2}
\eea

These equations derive from the potential :  

\beq
-\frac{\Delta c(g,h)}{24} = N \epsilon \ g^2  + (N-2)\epsilon\  h^2 - 2 \mathcal{D}_{SSA}\ g^2 \ h - \frac{2}{3} \mathcal{D}_{AAA} \  h^3 
\eeq

These are up to (including) the first non-trivial order of the perturbations in $g$ and $h$.

 The problem now amounts to justify the operator algebra expansions in (\ref{eqn9}) - (\ref{eqn11}) and to calculate the constants $\mathcal{D}_{SSA}$ and $\mathcal{D}_{AAA}$.

The efficient method for calculating the operator product expansions and defining the corresponding coefficients is the Coulomb gas formalism.

 Calculating directly the expansions of the products of the slightly relevant operators (\ref{relevantfields}) encounters a problem: the explicit form of the Coulomb gas representation for the $\z{N}{p}$ theory is not known. We shall get around this problem by using the coset representation for the $Z^{(2)}_{N}$ theory and the related techniques. In particular, we shall generalize the methode developed in papers \cite{ref11,ref12} for the $SU(2)$ cosets.

\section{Relating $\z{2r+1}{p} $ and the $\wb{r}{}$ theories }

The idea is to realize the parafermionic theory in terms of some simpler conformal theories. To do so we start with the coset representing $\z{N}{p}$ \cite{ref13}:


\beq
\z{N}{p}=\frac{SO_{k}(N)\otimes SO_{2}(N)}{SO_{k+2}(N)} \qquad p=N-2+k
\label{coset_ZN}
\eeq

Here $SO_{k}(N)$ is the orthogonal affine algebra of level $k$. This coset is a particular case of a symmetric coset $G_k \otimes G_l / G_{k+l}$, with a shift parameter $l=2$. Generally speaking, the higher the shift parameter, the more complex the theory. For instance the number of sectors is increasing with this shift parameter, as the chiral algebra becomes richer. Following \cite{ref11,ref12}, we decompose the coset with shift $l=2$ in terms of the several simpler $l=1$ cosets :


\beq
Z^{(2)}_{N}(p)\otimes\frac{SO_{1}(N)\otimes SO_{1}(N)}{SO_{2}(N)}=\frac{SO_{k}(N)\otimes SO_{1}(N)}{SO_{k+1}(N)}\otimes\frac{SO_{k+1}(N)\otimes SO_{1}(N)}{SO_{k+2}(N)}
\label{coset1}
\eeq

The two $l=1$ coset factors in the r.h.s., as well as the additional coset factor in the l.h.s., correspond to the $W\!B_{r}$ theories with rank $r=\frac{N-1}{2}$ : 

\beq
\wb{r}{(p)} = \frac{SO_{k}(2r+1)\otimes SO_{1}(2r+1)}{SO_{k+1}(2r+1)} \qquad p=2r - 1+ k 
\eeq

The equation (\ref{coset1}) reads :

\beq
Z^{(2)}_{N}(p)\otimes \wb{r}{(2r)}=\wb{r}{(p)}\otimes \wb{r}{(p+1)}
\label{coset2}
\eeq

This equation relates the representations of the corresponding algebras. It could be
reexpressed in terms of characters of representations, as is being usually done in the analyses of cosets. But this equation allows also to relate the conformal blocs of correlation functions. In doing so one relates the chiral (holomorphic) factors of physical operators. This later approach has been developped and analyzed in great detail in the papers \cite{ref11,ref12}, for the $SU(2)$ coset theories.

A $W\!B_{r}$ theory is a special case of $\mathcal{W}$ theories. They have been defined in \cite{ref14}. The $\wb{r}{p}$ chiral algebra is made of $r$ bosonic currents $W^{(2k)}$ with confomal dimension $2k$, $k=1, \dots r$ and one fermionic current $\Psi$ with dimension $r+1/2$, and its central charge is :

\beq
c_{\wb{r}{(p)}} = \left( r+ \frac{1}{2} \right)\left( 1 -  \frac{2r(2r-1)}{p(p+1)} \right)
\eeq


A direct consequence of (\ref{coset2}) is an egality for the central charges : 

\beq
c_{\z{N}{p}} + c_{\wb{r}{(2r)}} = c_{\wb{r}{(p)}} + c_{\wb{r}{(p+1)}}
\eeq 

\subsection{The $N=5$ case}

For the sake of simplicity we first treat the $N=5$ case.
The coset decomposition of $\z{5}{p}$ is : 

\beq
Z^{(2)}_{5}(p)\otimes \wb{2}{(p=4)} = \wb{2}{(p)} \otimes \wb{2}{(p+1)}
\label{cosetZ5}
\eeq

Our first step will be to identify $\z{5}{p}$ primary fields in $\wb{2}{(p)} \otimes \wb{2}{(p+1)}$. These fields $\Phi_{\left(\vec{n} | \vec{n}'\right)}$ are characterized by the $\z{5}{p}$ Kac formula which fixes their conformal dimensions \cite{ref6}:

\bea
\Delta_{\left(\vec{n} | \vec{n}'\right)} & = & \frac{\left((p+2)\vec{n} - p \vec{n}' \right)^2 - 10}{4p(p+2)} + B \\
n_1 + n_2 & < & p+2 \\
n_1' + n_2' & < & p
\eea

On the other hand, $\wb{2}{p}$ primary fields also obey a Kac formula : the primary field $\Phi_{\left(\vec{n} | \vec{n}'\right)}$ has conformal dimension \cite{ref14}:

\bea
\Delta_{\left(\vec{n} | \vec{n}'\right)} & = & \frac{\left((p+1)\vec{n} - p \vec{n}' \right)^2 - \frac{5}{2}}{2p(p+1)} + b \\
n_1 + n_2 & < & p+1 \\
n_1' + n_2' & < & p
\eea

When identifying fields between $ Z^{(2)}_{5}(p)\otimes \wb{2}{(p=4)}$ and $\wb{2}{(p)} \otimes \wb{2}{(p+1)} $, one obvious relation is the equality of the total conformal dimension. Using the Kac formulas of $\z{5}{p}$  and ${\wb{2}{(p)}}$, one can check the following identity : 

\beq 
\delta^{\wb{r}{(p)}}_{(\vec{n} \mid \vec{k})} + \delta^{\wb{r}{p+1}}_{(\vec{k} \mid \vec{n}')} = \delta^{\z{N}{p}}_{(\vec{n} \mid \vec{n}')}  + \left( \vec{k} - \frac{\vec{n}+\vec{n}'}{2}\right)^2
\label{dimeg} 
\eeq

Where $\delta_{(\vec{n}\mid \vec{n}')} =  \Delta_{(\vec{n}\mid \vec{n}')} - B $ is the dimension of $\Phi_{(\vec{n}\mid \vec{n}')}$ minus the boundary term : it corresponds to the Coulomb gas vertex operator part of the dimension. 

For operators, the coset relation (\ref{cosetZ5}) together with (\ref{dimeg}) motivates the following statement :

\beq
 \Phi^{(\z{5}{p})}_{(\vec{n}\mid \vec{n}')}  \otimes \Phi^{(\wb{2}{4})}_{(\vec{s}\mid \vec{s}')} 
=\sum_{\vec{k}} a_{\vec{k}} \Phi^{(\wb{2}{p})}_{(\vec{n}\mid \vec{k})}  \otimes
 \Phi^{(\wb{2}{p+1})}_{(\vec{k}\mid \vec{n}')} 
\label{decomp}
\eeq

The operators in this relation could be primaries or their descendants.

In other words, for any field from $\z{5}{p}$, there are fields in $\wb{2}{(p)}$, $\wb{2}{(p+1)}$ and $\wb{2}{(4)}$ such that the product of the fields as in (\ref{decomp}) have the same correlation functions.




From (\ref{decomp}) it appears that there exist some selection rules in $\wb{2}{(p)} \otimes \wb{2}{(p+1)}$ : only "diagonal" cross-products are pertinent in this analysis. By diagonal we mean product of the form $\Phi_{(\vec{n}\mid \vec{k})}\otimes \Phi_{(\vec{k'}\mid \vec{n}')}$ with $\vec{k}=\vec{k}'$. 
These features are discussed in much detail in the paper [12].

Equation (\ref{coset2}) should in fact read : 


\beq
\z{5}{p} \otimes \wb{2}{(4)} = P(\wb{2}{(p)} \otimes \wb{2}{(p+1)})
\label{coset3}
\eeq

where $P$ is a projector. In terms of primary fields, $P$ projects the product of all fields $\left\{ \phi^{(p)}_{(\vec{n} \mid \vec{k})} \otimes \phi^{(p+1)}_{(\vec{q} \mid \vec{n}')}  \right\} = \wb{r}{(p)} \otimes \wb{r}{(p+1)} $  to the subspace : 


\beq
P(\wb{2}{(p)} \otimes \wb{2}{(p+1)}) = \left\{ \phi^{(p)}_{(\vec{n} \mid \vec{k})} \otimes \phi^{(p+1)}_{(\vec{k} \mid \vec{n}')}  \right\} 
\label{projector}
\eeq

\subsubsection{The chiral algebra} 

It is interesting to take a closer look at the descendants of the identity, since they contain the chiral algebra of the theory. In particular one should be able to build the stress energy tensor of the second parafermionic theory in terms of fields living in $\wb{2}{p} \otimes \wb{2}{p+1}$. More precisely, equation (\ref{decomp}) enforces the following assumption : any chiral current $\Lambda(z)$ of $\z{5}{p}$ should have a decomposition of the following form :

\beq 
 \Lambda \otimes \Phi^{(p=4)} =  \sum_{\vec{k}} a_{\vec{k}}  \Phi^{(p)}_{(1,1 \mid \vec{k})}  \otimes
 \Phi^{(p+1)}_{(\vec{k}\mid 1,1)} 
\eeq

where the fields involved can be either primary fields or their descendants.

At level $\Delta = 0 $, we get the trivial 
\beq \mathbb{I}^{\z{5}{p}} \otimes \mathbb{I}^{(p=4)} = \mathbb{I}^{(p)} \otimes \mathbb{I}^{(p+1)} 
\eeq

At level $\Delta = 1$, there is not much liberty either : we have only one field of conformal dimension 1 in both $Z^{(2)}_{5}(p)\otimes \wb{2}{(p=4)}$ and $  \wb{2}{(p)} \otimes \wb{2}{(p+1)} $ : 

\beq 
\mathbb{I}^{\z{5}{p}} \otimes \Phi^{(p=4)}_{ (1,1  \mid 3,1)} = \Phi^{(p)}_{ (1,1  \mid 2,1)} \otimes \Phi^{(p+1)}_{ (2,1  \mid 1,1)}
\eeq

Things get more interesting for $\Delta = 2$. One can prove the following decompositions :

\bea
T^{(p=4)} & = & \frac{p}{5(p+4)} T^{(p)} + \frac{p+2}{5(p-2)} T^{(p+1)} + \sqrt{\frac{2(p+5)(p-3)}{5(p+4)(p-2)}} \Phi^{(p)}_{ (1,1  \mid 1,3)} \otimes \Phi^{(p+1)}_{ (3,1  \mid 1,1)} \\
T^{\z{5}{p}} & = & \frac{4}{5}\frac{p+5}{(p+4)} T^{(p)} + \frac{4}{5}\frac{p-3}{(p-2)} T^{(p+1)} - \sqrt{\frac{2(p+5)(p-3)}{5(p+4)(p-2)}} \Phi^{(p)}_{ (1,1  \mid 1,3)} \otimes \Phi^{(p+1)}_{ (3,1  \mid 1,1)}
\eea

One can check that these fields obey the required OPEs : 

\bea
T^{(p=4)}\ (z)  \ T^{(p=4)}\ (0) & =  & \frac{1/2}{z^4} + \frac{2 T^{(p=4)}}{z^2} + \frac{\partial T^{(p=4)}}{z} + \mathcal{O}(1) \\
T^{\z{5}{p}}\ (z)  \ T^{\z{5}{p}}\ (0) & =  & \frac{c(5,p)/2}{z^4} + \frac{2 T^{\z{5}{p}}}{z^2} + \frac{\partial T^{\z{5}{p}}}{z} + \mathcal{O}(1) \\
T^{\z{5}{p}}\  (z)  \ T^{(p=4)}\ (0) & = & \mathcal{O}(1)
\eea

A few remarks are in order ar this point : we are dealing with the holomorphic part of the fields only. So when doing the expansions of the products of $\wb{2}{(p)}$ and $\wb{2}{(p+1)}$ operators, one does them :

1) with the square roots of the constants;

2) one keeps in these expansions the "diagonal" cross-products only: the products of $\wb{2}{(p)}$ and $\wb{2}{(p+1)}$ operators which appear in coset relations for operators, 
due to eq.(\ref{coset3}).

3) one needs to know some $\wb{2}{p}$ algebra constants, for which the Coulomb gas is known (cf part 3) 

\subsubsection{Singlets}

Neutral primary fields in $\z{5}{p}$ are referred to as singlets. They belong to the simplest sector of the $N=5$ second parafermionic theory, and they enjoy a zero boundary term in their conformal dimensions. $\Phi_{\left(\vec{n} | \vec{n}'\right)}$ is a singlet when $n_1' -n_1 = 0$ mod $2$ and $n_2'-n_2 = 0$ mod $4$. We will denote such a field  $S_{\left(\vec{n} | \vec{n}'\right)}$. We remark that $\vec{k} = \frac{\vec{n}+\vec{n}'}{2}$ is then an admissible weight of $B_2$.

Equations (\ref{dimeg}) and (\ref{decomp}) lead to :

\beq 
S^{(p)}_{(\vec{n}\mid \vec{n}')} \otimes  \mathbb{I}= \Phi^{(p)}_{\left(\vec{n}\mid \frac{\vec{n}+\vec{n}'}{2}\right)} \otimes \Phi^{(p+1)}_{\left(\frac{\vec{n}+\vec{n}'}{2}\mid \vec{n}'\right)}
\label{singlet}
\eeq



In the l.h.s of (\ref{singlet}) the field in $\wb{2}{(p=4)}$ is the identity $\mathbb{I}$ so it can be dropped. The notation $\Phi^{(p)}$ stands for fields living in $\wb{2}{(p)}$.

In general neutral operators are expected to have a simple decomposition. In particular they will have a trivial $\mathbb{I}$ term in $\wb{2}{(4)}$, allowing to express them in $\wb{2}{(p)} \otimes \wb{2}{(p+1)}$ only. The fields we will use to perturbate $\z{N}{p}$ are neutral since we want to keep the $Z_N$ symmetry.

For instance we have the following identification :

\beq
S = \mathcal{S}^{\z{5}{p}}_{(1 1 \mid 3 1 )} = \Phi^{(p)}_{(1 1 \mid 2 1 )} \otimes \Phi^{(p+1)}_{(2 1  \mid 3 1 )}
\label{decompS}
\eeq


\subsubsection{Doublets}

Doublets $\mathcal{D}^{Q}$ are charged primary fields of $\z{5}{p}$ w.r.t. the $Z_{5}$ symmetry. They belong to a less trivial sector than singlets, and they have a non trivial boundary term in their dimension. Therefore their decomposition in (\ref{decomp}) requires a non trivial field in $\wb{2}{(p=4)}$, to make up for the missing boundary terms in equation (\ref{dimeg}).

But the $\z{5}{p}$ fields we use to perturb with have to be neutral to conserve the $Z_5$ symmetry. Thus we are interested in neutral descendants of doublets, like $\psi^{\pm Q}_{-\delta_Q +n} \mathcal{D}^{\mp Q} $ with fundamental gap $\delta_{Q} = \frac{Q(Q-N)}{2N}$ mod[1].

Taking into account both the boundary term and the descendant gap, equations (\ref{dimeg}) and (\ref{decomp}) give :

\beq
\psi^{\pm Q}_{-\delta_Q }\mathcal{D}^{\mp Q}_{(\vec{n}\mid \vec{n}')} \otimes \mathbb{I}^{(p=4)} = \sum_{\vec{k}} a(\vec{k}) \Phi^{(\wb{2}{p})}_{(\vec{n}\mid \vec{k})}\otimes
\Phi^{(\wb{2}{p+1})}_{(\vec{k}\mid \vec{n}')}
\label{doublet}
\eeq
where the sum can be restricted, using (\ref{dimeg}), to $\vec{k}$ obeying : $\left( \vec{k} - \frac{\vec{n}+\vec{n}'}{2}\right)^2 = B_Q + \delta_Q$, so that the sum is actually finite.

Let's take an exemple : the doublet $\mathcal{D}^{q= \pm 1}_{(1 1  \mid 1 3)}$.

This is one of the 2 fundamental $q=1$ ($Q=2$) doublets in $\z{5}{p}$ . The structure of its module is such that there is only one neutral descendant with gap $\delta_Q= \frac{2}{5}$, because one of the degeneracy condition reads $\psi^{-1}_{-\frac{2}{5}} \mathcal{D}^{1}_{(1 1  \mid 1 3)} = \psi^{1}_{-\frac{2}{5}} \mathcal{D}^{-1}_{(1 1  \mid 1 3)}$ \cite{ref6}. Applying (\ref{doublet}) here gives :

\beq
\psi_{-\frac{2}{5}} \mathcal{D}^{-1}_{(1 1   \mid 1 3)} = \sum_{\vec{k}} a_{\vec{k}} \Phi^{(\wb{2}{p})}_{(1 1  \mid \vec{k})}\otimes
\Phi^{(\wb{2}{p+1})}_{(\vec{k}\mid 1 3)} 
\label{decomp1} 
\eeq

Here the sum occurs for $ \left( \vec{k} - (1,2)\right)^2 = B_Q + \delta_Q = \frac{1}{2} $, whose solutions are :

\beq
\vec{k} = \left\{ \begin{array}{l}
(1,1)\\
(1,3)\\
(2,1)
\end{array} \right. \label{resultk1} \eeq

We note here that the fields in $\wb{r}{}$ are of two sorts : Neveu-Schwartz (if $n_r-n_r'$ is even) and Ramond (if $n_r-n_r'$ is odd). Ramond field have a boudary term in their conformal dimension : $B_r = \frac{1}{16}$, therefore equation (\ref{dimeg}) exclude Ramond fields when we decompose a neutral field of $\z{N}{p}$. So that $\vec{k}$ must also obey $k_r = 1$ mod 2.

(\ref{decomp1}) and (\ref{resultk1}) sum up to : 

\beq
\psi_{-\frac{2}{5}} \mathcal{D}^{-1}_{(1 1  \mid 1 3)}  =  a\  \Phi^{(\wb{2}{p})}_{(1 1  \mid 1 1 )}\otimes
\Phi^{(\wb{2}{p+1})}_{(1 1 \mid 1 3)}   +  b\  \Phi^{(\wb{2}{p})}_{(1 1 \mid 1 3)}\otimes
\Phi^{(\wb{2}{p+1})}_{(1 3 \mid 1 3)}   +  c\  \Phi^{(\wb{2}{p})}_{(1 1 \mid 2 1)}\otimes
\Phi^{(\wb{2}{p+1})}_{(2 1 \mid 1 3)}
\label{decompA}
\eeq

The coefficient $a,b,c$ are still to be determined at this point. Several methods can be used to calculate them. One can use the expression of the stress-energy tensor $T^{(p=4)}$ and demand that the field $\psi_{-\frac{2}{5}} \mathcal{D}^{-1}_{(1 1  \mid 1 3)} \otimes \mathbb{I}^{(p=4)}$ has the right conformal dimensions w.r.t. $\wb{2}{(p=4)}$, i.e. $\Delta^{(p=4)} =0$. One other way to determine these constants is through the fusion rules of $\z{5}{p}$ : imposing the fusion rule $A \times A \rightarrow A$, or $S \times S \rightarrow A$, will fix  $a,b,c$ uniquely :

\bea 
S(z)\times S(0)  & \rightarrow & \frac{\sqrt{\mathcal{D}_{S S A}}}{z^{2\Delta_S - \Delta_A}}  \left\{ A(0) + \mathcal{O}(z)   \right\} \label{fusionSSA}\\ 
A(z) \times A(0) & \rightarrow &\frac{1}{z^{2\Delta_A}} \left\{ \mathbb{I} + \mathcal{O}(z) \right\}
\label{fusionAAI}
\eea
Together with (\ref{decompS}), (\ref{fusionSSA}) and (\ref{fusionAAI}) allow to express the coefficients $a,b,c$ of equation (\ref{decompA}) in terms of algebra constants of $\wb{2}{}$.

Injecting (\ref{decompA}) in (\ref{fusionAAI}) gives : 

\beq
a^2+b^2+c^2  =  1
\eeq

Putting (\ref{decompA},\ref{decompS}) in (\ref{fusionSSA}) gives : 

\bea
\sqrt{D^{(p+1)}_{(2 1  \mid 3 1 )(2 1  \mid 3 1 )(1 1  \mid 1 3)}} & = & a\sqrt{\mathcal{D}_{SSA}} \label{eqa} \\ 
\sqrt{D^{(p)}_{(1 1  \mid 2 1 )(1 1  \mid 2 1 )(1 1  \mid 1 3)}D^{(p+1)}_{(2 1  \mid 3 1 )(2 1  \mid 3 1 )(1 3\mid 1 3)}} & = & b\sqrt{\mathcal{D}_{SSA}} \label{eqb} \\
\sqrt{D^{(p)}_{(1 1  \mid 2 1 )(1 1  \mid 2 1 )(1 1  \mid 2 1 )}D^{(p+1)}_{(2 1  \mid 3 1 )(2 1  \mid 3 1 )(2 1  \mid 1 3)}} & = & c\sqrt{\mathcal{D}_{SSA}} \label{eqc}
\eea

For clarity we have adopted the following notations :  $D_{(\ldots)}^{(p)}$ stands for a fusion constant of $\wb{2}{(p)}$, while $\mathcal{D}_{(\ldots)}^{(p)}$ corresponds to a $\z{5}{p}$ constant. 

Knowing the $\wb{2}{}$ algebra constants $D^{(\ldots)}_{(\ldots)}$ then allows to determine $a,b,c$ and then $\mathcal{D}_{SSA}$ and $\mathcal{D}_{AAA}$. The problem of evaluating these $\z{5}{p}$ algebra constants has been reduced to the calculation of some $\wb{2}{(p)}$ algebra constants.

As it was said above, the chiral factor operators are related to the conformal bloc functions, not to the actual physical correlators. On the other hand, the coefficients of the operator algebra expansions are defined by the three point functions. These latter are factorizable, into holomorphic - antiholomorphic functions. So that, when the relation is established on the level of chiral factor operators, for the holomorphic three point functions, this relation could then be easily lifted to the relation for the physical correlation functions. Saying it differently, with the relations for the chiral factor operators one should be able to define the square roots of the physical operator algebra constants.

In the part 3 we will calculate those $\wb{2}{}$ constants we need, to obtain  for $A$ and $S$ the following decomposition at leading order in $\epsilon$ : 

\beq \begin{array}{lll} 
S & = & \Phi^{(p)}_{(1 1 \mid 2 1  )} \otimes \Phi^{(p+1)}_{(2 1  \mid 3 1  )}\\
A & = & \frac{1}{\sqrt{2}} \Phi^{(p)}_{(1 1  \mid 1 1 )}\otimes \Phi^{(p+1)}_{(1 1  \mid 1 3 )} + \frac{1}{\sqrt{2}} \Phi^{(p)}_{(1 1  \mid 1 3)}\otimes \Phi^{(p+1)}_{(1 3  \mid 1 3 )}
\end{array}
\eeq





\subsection{The case $N \geq 7$}

This construction can be generalized to the case $N=2r+1$ with $r \geq 3$.
For the $\z{2r+1}{p}$ parafermionic theory the coset relation reads : 

\beq
\z{2r+1}{p} \otimes \wb{r}{(2r)} = P(\wb{r}{(p)} \otimes \wb{r}{(p+1)})
\label{cosetN}
\eeq

We recall that the two $\z{2r+1}{p}$ slightly relevant fields $S$ and $A$ are :

\beq \begin{array}{lll} 
S & = & \mathcal{S}_{(1 1 1 \ldots   \mid 3 1 1 \ldots  )} \\
A & = & \psi^{-1}_{-\frac{2}{N}} \mathcal{D}_{(1 1 1 \ldots \mid 1 2 1 \ldots)}
\end{array}
\eeq

Using the same method as for the $N=5$ case we obtain  :

\beq \begin{array}{lll} 
S & = & \Phi^{(p)}_{(1 1 1 \ldots   \mid 2 1 1 \ldots  )} \otimes \Phi^{(p+1)}_{(2 1 1 \ldots   \mid 3 1 1 \ldots  )}\\
A & = & \frac{1}{\sqrt{2}} \Phi^{(p)}_{(1 1 1 \ldots \mid 1 1 1 \ldots)}\otimes \Phi^{(p+1)}_{(1 1 1 \ldots   \mid 1 2 1 \ldots  )} + \frac{1}{\sqrt{2}} \Phi^{(p)}_{(1 1 1 \ldots \mid 1 2 1 \ldots)}\otimes \Phi^{(p+1)}_{(1 2 1 \ldots   \mid 1 2 1 \ldots  )}
\end{array}
\eeq

\subsection{Summary}

We want to perturbe $\z{N}{p}$ with the fields $A$ and $S$. We need to know the algebra constants $\mathcal{D}_{SSS}$,  $\mathcal{D}_{SSA}$, $\mathcal{D}_{SAA}$ and $\mathcal{D}_{AAA}$. Here are the relations obtained with the coset construction (\ref{coset1}) : 

\bea
\sqrt{\mathcal{D}_{SSS}} & = &  \sqrt{D^{(p)}_{(1 1 1 \ldots \mid 2 1 1 \ldots)(1 1 1 \ldots \mid 2 1 1 \ldots)(1 1 1 \ldots \mid 2 1 1 \ldots)}D^{(p+1)}_{(2 1 1 \ldots \mid 3 1 1 \ldots)(2 1 1 \ldots \mid 3 1 1 \ldots)(2 1 1 \ldots \mid 3 1 1 \ldots)}} \label{DSSS}\\
\sqrt{\mathcal{D}_{SSA}} & = &  a^{-1}\sqrt{D^{(p+1)}_{(2 1 1 \ldots \mid 3 1 1 \ldots)(2 1 1 \ldots \mid 3 1 1 \ldots)(1 1 1 \ldots \mid 1 2 1 \ldots)}} \\ 
\sqrt{\mathcal{D}_{SAA}} & = & \frac{b}{a} \sqrt{D^{(p+1)}_{(2 1 1 \ldots \mid 3 1 1 \ldots)(1 2 1 \ldots \mid 1 2 1 \ldots)(1 2 1 \ldots \mid 1 2 1 \ldots)}} \label{DAAS}\\
\sqrt{\mathcal{D}_{AAA}} & = & a \sqrt{D^{(p+1)}_{(1 1 1 \ldots \mid 1 2 1 \ldots)(1 1 1 \ldots \mid 1 2 1 \ldots)(1 1 1 \ldots \mid 1 2 1 \ldots)}} \\ & & + \frac{b^2}{a} \sqrt{D^{(p+1)}_{(1 2 1 \ldots \mid 1 2 1 \ldots)(1 2 1 \ldots \mid 1 2 1 \ldots)(1 1 1 \ldots \mid 1 2 1 \ldots)}} \\
 & & + \frac{c^2}{a} \sqrt{D^{(p+1)}_{(2 1 1 \ldots \mid 3 1 1 \ldots)(2 1 1 \ldots \mid 3 1 1 \ldots)(1 1 1 \ldots \mid 1 2 1 \ldots)}} 
\eea

\section{Calculation of the $\wb{r}{}$ algebra constants}


The $\wb{r}{}$ Coulomb Gas is known \cite{ref14}, therefore we have integral representations of the fusion algebra constants. Unfortunately we don't know how to calculate the most general form of these integrals. We will show in this part how to obtain the constants we need.

\subsection{The $\wb{r}{}$ Coulomb Gas}

We need to calculate some fusion algebra constants of $\wb{r}{(p)}$. 
For these theories the Coulomb gas representation is made of $r$ bosonic fields ${\varphi}_i$, quantized with a background charge and the Ising model fields: $\Psi$ (free fermion) and $\sigma$ (spin operator) \cite{ref14}.

The details about the $\wb{r}{}$ Coulomb gas are given in appendix \ref{appendixCG}. Three point functions have the following form : 

\bea 
\left< V_1(0) V_2(1) \overline{V}_3(\infty)   \right> =\left< V_1(0) V_2(1) \prod_{a=1}^{r} \left(\frac{1}{k^+_a!}\prod_{i=1}^{k_a^+}  \int \ud^2u_i^{(a)} V_a^{+}(u_i^{(a)},\bar{u}_i^{(a)})\right) \right. \cr
\left. \left(\frac{1}{k^-_a!} \prod_{j=1}^{k_a^-}  \int  \ud^2v_j^{(a)}  V_a^{-}(v_j^{(a)},\bar{v}_j^{(a)}) \right) \overline{V}_3 (\infty) \right> 
\label{gen_integral1}
\eea

where $k_a^{\pm}$ are the number of screening operators $V_{a}^{\pm}$ required to ensure the neutrality condition : 

\beq 
\begin{array}{lll}
\sum_a k_a^{+} \vec{e}_{a}  & = & \sum_i \left(n_i^{1} + n_i^{2} - n_i^{3} - 1\right)\vec{\omega}_i\\
\sum_a k_a^{-} \vec{e}_{a}  & = & \sum_i \left(n_i'^{1} + n_i'^{2} - n_i'^{3} - 1\right)\vec{\omega}_i
\end{array}
\eeq

As usual in the Coulomb Gas approach, the vertex operators representing the primary fields have non trivial normalizations. 

We will denote as $C_{a \ b}^{c}$ the fusion constants obtained in the Coulomb Gaz representation (i.e. 3 point functions) and $D_{a\ b\ c}$ the actual $\wb{r}{}$ constants : 

\bea
C_{a \ b}^{c} & = &\left< V_a(0) V_b(1) \overline{V}_c(\infty)   \right> \\
D_{a\ b\ c} & = &\left< \phi_a(0) \phi_b(1) \phi_c(\infty)   \right> 
\eea 

These two quantities are related by :

\beq
C_{a \ b}^{c} = N_a N_b N_c^{-1} D_{a\ b\ c}
\label{sym}
\eeq

$D_{a\ b\ c}$ being symetric under any permutation of $a,b,c$ and $N_a$ is the normalization of the vertex $V_a$ : 

\beq
N_a^2 = C_{a \ a}^{\mathbb{I}} = \left< V_a(0) V_a(1) \right>
\label{N2}
\eeq

So that 

\bea 
N_{\phi}^2 =\left< V_{\phi}(0) V_{\phi}(1) \prod_{a=1}^{r} \left(\frac{1}{k^+_a!}\prod_{i=1}^{k_a^+}  \int \ud^2u_i^{(a)} V_a^{+}(u_i^{(a)},\bar{u}_i^{(a)})\right) \right. \cr
\left. \left(\frac{1}{k^-_a!}\prod_{j=1}^{k_a^-}  \int  \ud^2v_j^{(a)}  V_a^{-}(v_j^{(a)},\bar{v}_j^{(a)}) \right) V_{2\vec{\alpha}_0} (\infty) \right> 
\label{gen_integral2}
\eea

Unfortunately we don't know how to calculate these integrals in the general case.

\subsection{Some easy integrals}

Evaluating the general form of integrals (\ref{gen_integral2}) could prove quite involved. Luckily we are interested in algebra constants involving fields with relatively small indices, so that the number of screening operators should remain reasonable. Furthermore, since one of the screening operator is fermionic, we can already predict the vanishing of some integrals : whenever a three point function requires an odd number of fermionic screening operators, the corresponding constant will obviously be zero (at least in the Neveu-Schwarz sector).

This is the case for the following $\wb{r}{}$ constants : 

\bea
\forall \ \  (\vec{n},\vec{m}) &  &  \quad D_{ (\vec{n} \mid \vec{m}) \   ( \vec{n} \mid \vec{m} ) \  (1 1 1 \ldots \mid 2 1 1 \ldots)} = 0 \\
& & \quad D_{ (\vec{n} \mid \vec{m}) \   ( \vec{n} \mid \vec{m} ) \  (2 1 1 \ldots \mid 3 1 1 \ldots)} = 0
\eea



Going back to the fields $A$ and $S$ in $\z{N}{p}$, this implies the following trivial results:

\bea
\mathcal{D}_{AAS} & = & 0 \\
\mathcal{D}_{SSS} & = & 0
\eea

\subsection{Some other integrals} 
\label{int_doable}

Integrals involving only a few screening operators can be calculated exactly. This is the case for the following algebra constants :

\beq
\forall \ \  (\vec{n},\vec{m}) \quad C_{ (\vec{n} \mid \vec{m}) \     (1 1 1 \ldots \mid 1 2 1 \ldots)}^{( \vec{n} \mid \vec{m} )} 
\eeq






For all these constants the neutrality condition reads : 

\beq
\begin{array}{lll}
\sum_a k_a^{+} \vec{e}_{a}  & = & 0 \\
\sum_a k_a^{-} \vec{e}_{a}  & = & \vec{\omega}_2
\end{array}
\eeq

which fixes the number of screening operators :

\beq
\begin{array}{lll}
k^{+} & = & (0, 0, 0, \dots ,0)\\
k^{-} & = & (1, 2, 2, \dots ,2)
\end{array}
\eeq

For instance let us caculate $C_{(1 1 1 \ldots \mid 1 2 1 \ldots)(1 1 1 \ldots \mid 1 2 1 \ldots)}^{(1 1 1 \ldots \mid 1 2 1 \ldots)}$ :

\bea
C_{(1 1 1 \ldots \mid 1 2 1 \ldots)(1 1 1 \ldots \mid 1 2 1 \ldots)}^{(1 1 1 \ldots \mid 1 2 1 \ldots)} = \frac{1}{(2!)^{r-1}}   \left<       \  	\mathrm{exp}\left(-i\alpha_{-}\frac{\vec{\omega}_2}{\sqrt{2}}.\vec{\varphi}(0)\right) \mathrm{exp} \left( -i\alpha_{-} \frac{\vec{\omega}_2}{\sqrt{2}} . \vec{\varphi}(0)\right) \right.\cr
\int \ud^2 u^{(1)}_1 \mathrm{exp}\left(i\alpha_{-}\frac{\vec{e}_1}{\sqrt{2}}.\vec{\varphi}(u^{(1)}_1,\bar{u}^{(1)}_{1})\right)  
\prod_{2 \leq a \leq r-1} \prod_{i=1,2} \int \ud^2u^{(a)}_{i} \mathrm{exp} \left( i\alpha_{-} \frac{\vec{e}_a}{\sqrt{2}} . \vec{\varphi} (u^{(a)}_i,\bar{u}^{(a)}_{i}) \right) \cr
 \left. \prod_{i=1,2} \int \ud^2u^{(r)}_{i}\psi(u^{(r)}_i)\overline{\psi}(\bar{u}^{(r)}_i) \mathrm{exp} \left( i\alpha_{-} \frac{\vec{e}_r}{\sqrt{2}} .\vec{\varphi}(u^{(r)}_i,\bar{u}^{(r)}_{i})\right) \mathrm{exp}\left( \left(2\vec{\alpha}_0 +i\alpha_{-} \frac{\vec{\omega}_2}{\sqrt{2}} \right) . \vec{\varphi} (\infty)\right) \right>
\eea

This integral can be evaluated by first integrating over $\left( u^{(r)}_1,u^{(r)}_2 \right)$, then over $\left( u^{(r-1)}_1,u^{(r-1)}_2 \right)$, etc..

Proceeding in this fashion we find the following results ( we give only the leading order in $\epsilon = \frac{1}{p}$ because that is all we need for the renormalization group method ):

\bea
C_{(1 1 1 \ldots \mid 1 2 1 \ldots)(1 1 1 \ldots \mid 1 2 1 \ldots)}^{(1 1 1 \ldots \mid 1 2 1 \ldots)} & = &  2(2r-1) \left(\frac{\pi}{\epsilon}\right)^{2r-1}  \\ 
C_{(1 2 1 \ldots \mid 1 2 1 \ldots)(1 1 1 \ldots \mid 1 2 1 \ldots)}^{(1 2 1 \ldots \mid 1 2 1 \ldots)} & = & 2(2r-1) \left(\frac{\pi}{\epsilon}\right)^{2r-1} \epsilon^2 \\ 
C_{(2 1 1 \ldots \mid 3 1 1 \ldots)(1 1 1 \ldots \mid 1 2 1 \ldots)}^{(2 1 1 \ldots \mid 3 1 1 \ldots)} & = &  \frac{2r+1}{2} \left(\frac{\pi}{\epsilon}\right)^{2r-1}  \\
C_{(1 1 1 \ldots \mid 2 1 1 \ldots)(1 1 1 \ldots \mid 1 2 1 \ldots)}^{(1 1 1 \ldots \mid 2 1 1 \ldots)} & = & 2r \left(\frac{\pi}{\epsilon}\right)^{2r-1}  
\eea

\subsection{Some more involved integrals}

On the other hand, the calculation of $C_{(1 2 1 \ldots \mid 1 2 1 \ldots)(1 2 1 \ldots \mid 1 2 1 \ldots)}^{(1 2 1 \ldots \mid 1 2 1 \ldots)}$, $C_{(2 1 1 \ldots \mid 3 1 1 \ldots)(1 2 1 \ldots \mid 1 2 1 \ldots)}^{(2 1 1 \ldots \mid 3 1 1 \ldots)}$ is a bit more involved. Because the number of screening operators is twice as much, the same method won't work. 
\beq
\begin{array}{lll}
k^{+} & = & (1, 2, 2, \dots )\\
k^{-} & = & (1, 2, 2, \dots )
\end{array}
\eeq

Instead we will calculate the following 4 points correlation fuction, and use it to derive a simpler expression for these constants : 

\beq
f(z,\bar{z}) = \left< \phi_a(0) \phi_{(1 1 1 \ldots\mid 1 2 1 \ldots )}(z,\bar{z}) \phi_{(1 2 1 \ldots \mid 1 1 1 \ldots )}(1) \phi_a(\infty)  \right>
\eeq
$\phi_a$ being an arbitrary field.

This function is single-channeled, therefore it factorizes : $f(z,\bar{z}) = f(z)\bar{f}(\bar{z})$  
\beq
f(z) = \frac{P(z)}{z^{\Delta_{(1 1 1 \ldots\mid 1 2 1 \ldots )}}(1-z)^2} 
\eeq
$P(z)=a_0 + a_1 z + a_2 z^2$ being a polynom of degree 2 whose coefficients are fixed by the fusion rules :
\bea
\phi_{(1 1 1 \ldots\mid 1 2 1 \ldots )} \times \phi_{(1 2 1 \ldots\mid 1 1 1 \ldots )} & \rightarrow & \sqrt{D_{(1 1 1 \ldots\mid 1 2 1 \ldots ),(1 2 1 \ldots\mid 1 1 1 \ldots ),(1 2 1 \ldots\mid 1 2 1 \ldots )}} \phi_{(1 2 1 \ldots\mid 1 2 1 \ldots )} \\
\phi_{(1 1 1 \ldots\mid 1 2 1 \ldots )} \times \phi_a & \rightarrow & \sqrt{D_{a,a,(1 1 1 \ldots\mid 1 2 1 \ldots )}} \phi_a \label{fusionphiaa}\\ 
\phi_{(1 2 1 \ldots\mid 1 2 1 \ldots )} \times \phi_a & \rightarrow & \sqrt{D_{a,a,(1 2 1 \ldots\mid 1 2 1 \ldots )}} \phi_a
\eea

We find that : 
\bea
a_0 = a_2 &  = &\sqrt{D_{a,a,(1 1 1 \ldots\mid 1 2 1 \ldots )}D_{a,a,(1 2 1 \ldots\mid 1 1 1 \ldots )}} \\
P(1) & = & \sqrt{D_{(1 1 1 \ldots\mid 1 2 1 \ldots ),(1 2 1 \ldots\mid 1 1 1 \ldots ),(1 2 1 \ldots\mid 1 2 1 \ldots )}D_{a,a,(1 2 1 \ldots\mid 1 2 1 \ldots )}}
\eea

An important point here is that for $W$ theories the modes $W^{2n}_{-1}$ are proportionnal to $L_{-1} =\partial$. So that the only descendant at level 1 of any pimary field $\Phi$ is just $\partial \Phi$. Thus we can write the next term in the fusion of $\phi_{(1 1 1 \ldots \mid 1 2 1 \ldots )}$ with $\phi_a$ : 

\beq
\phi_{(1 1 1 \ldots \mid 1 2 1 \ldots )}(z) \times \phi_a(0)  \rightarrow  \frac{\sqrt{D_{(1 1 1 \ldots\mid 1 2 1 \ldots ),(1 2 1 \ldots\mid 1 1 1 \ldots ),(1 2 1 \ldots\mid 1 2 1 \ldots )}}}{z^{\Delta_{(1 1 1 \ldots\mid 1 2 1 \ldots )}}} \left( \phi_a(0) + \beta^{-1} z \partial \phi_a(0) + \mathcal{O}(z^2) \right) 
\eeq
where $\beta^{-1}$ is fixed by conformal invariance alone: 

\beq
\beta^{-1} = \frac{\Delta_{(1 1 1 \ldots\mid 1 2 1 \ldots )}}{2\Delta_a}
\eeq

That way we have the additionnal relation : 

\beq
P(1) =  a_0 \frac{\Delta_{(1 1 1 \ldots\mid 1 2 1 \ldots )}\Delta_{(1 2 1 \ldots\mid 1 1 1 \ldots )}}{2\Delta_a}
\eeq

which translates into : 

\beq
D_{a,a,(1 2 1 \ldots\mid 1 2 1 \ldots )} = \left(\frac{\Delta_{(1 1 1 \ldots\mid 1 2 1 \ldots )}\Delta_{(1 2 1 \ldots\mid 1 1 1 \ldots )}}{2\Delta_a}\right) ^2 \frac{D_{a,a,(1 1 1 \ldots\mid 1 2 1 \ldots )}D_{a,a,(1 2 1 \ldots\mid 1 1 1 \ldots )}}{D_{(1 1 1 \ldots\mid 1 2 1 \ldots ),(1 2 1 \ldots\mid 1 1 1 \ldots ),(1 2 1 \ldots\mid 1 2 1 \ldots )}}
\eeq

Going back to the Coulomb gas, this involve the following constants :

\beq
\begin{array}{l}
C_{a\ (1 1 1 \ldots\mid 1 2 1 \ldots )}^{a}\\
C_{a\ (1 2 1 \ldots\mid 1 1 1 \ldots )}^{a}
\end{array}
\eeq

which we know how to calculate (cf part \ref{int_doable})
 
and the trivial $C_{(1 1 1 \ldots\mid 1 2 1 \ldots )(1 2 1 \ldots\mid 1 1 1 \ldots )}^{(1 2 1 \ldots\mid 1 2 1 \ldots )} = 1$, since it involves no screening operator. 

\subsection{The normalization integrals}

The only quantity which remains now is the normalization of the Coulomb gas vertex operators. Trying to evaluate directly (\ref{gen_integral2}) will encounter the same kind of problems we just had : too many screening operators are involved. 

Alternatively, recalling the cyclic symmetry of $D_{a\ b\ c}$ , one finds : 
\bea
C_{a\ b}^{c} & = & N_a N_b N_c^{-1} D_{a\ b\ c} \label{num}\\
C_{a\ c}^{b} & = & N_a N_c N_b^{-1} D_{a\ b\ c} \label{denom}
\eea

This leads to the following identity : 
\beq
\left(\frac{N_b}{N_c}\right)^2 = \frac{C_{a\ b}^{c} }{C_{a\ c}^{b} }
\label{NN}
\eeq

For instance $N^2_{(1 1 1 \ldots\mid 1 2 1 \ldots )} = C_{(1 1 1 \ldots\mid 1 2 1 \ldots )(1 1 1 \ldots\mid 1 2 1 \ldots )}^{(1 1 1 \ldots\mid 1 1 1 \ldots )}$ cannot be evalulated directly. But (\ref{NN}) gives us the following expression expression : 
\beq
\forall \  (a,c) \quad \left(\frac{N_{(1 1 1 \ldots\mid 1 2 1 \ldots )}}{N_{c}}\right)^2 = \frac{C_{a\ (1 1 1 \ldots\mid 1 2 1 \ldots )}^{c} }{C_{a\ c}^{(1 1 1 \ldots\mid 1 2 1 \ldots )} } 
\eeq
 
Now, choosing carefully the fields $a$ and $c$ makes the calculation possible : for instance one can take $a=c=(1 1 1 \ldots\mid 2 1 1 \ldots )$. The constraints, when choosing these fields, are the following : 

\begin{itemize}
\item one has to be able to evaluate $C_{a\ b}^{c}$ and $ C_{a\ c}^{b}$ 
\item one has to be able to calculate $N_c$
\item $C_{a\ b}^{c} \neq 0$ 
\end{itemize}

We obtain finally : 
\beq 
N_{(1 1 1 \ldots\mid 1 2 1 \ldots )}^2 = \frac{C_{(1 1 1 \ldots\mid 2 1 1 \ldots )\ (1 1 1 \ldots\mid 1 2 1 \ldots )}^{(1 1 1 \ldots\mid 2 1 1 \ldots )} }{C_{(1 1 1 \ldots\mid 2 1 1 \ldots )\ (1 1 1 \ldots\mid 2 1 1 \ldots )}^{(1 1 1 \ldots\mid 1 2 1 \ldots )} } C_{(1 1 1 \ldots\mid 2 1 1 \ldots )\ (1 1 1 \ldots\mid 2 1 1 \ldots )}^{(1 1 1 \ldots\mid 1 1 1 \ldots )} 
\eeq
All the constants appearing here are then evaluated as in section \ref{int_doable}. That way we find :
\beq
N_{(1 1 1 \ldots\mid 1 2 1 \ldots )}^2 = r(2r+1) \left(  \frac{\pi}{\epsilon} \right)^{2(2r-1)} \left( 1 + \mathcal{O}(\epsilon) \right)
\eeq

Generalizing this method allows one to evaluate all normalizations we need. For instance : 
\bea
\left(\frac{N_{(1 2 1 \ldots\mid 1 2 1 \ldots )}}{N_{(1 1 1 \ldots\mid 2 1 1 \ldots )}}\right)^2  & =  & \frac{C_{(2 1 1 \ldots\mid 1 1 1 \ldots )\ (1 2 1 \ldots\mid 1 2 1 \ldots )}^{(2 1 1 \ldots\mid 1 2 1 \ldots )} }{C_{(2 1 1 \ldots\mid 1 1 1 \ldots )\ (2 1 1 \ldots\mid 1 2 1 \ldots )}^{(1 2 1 \ldots\mid 1 2 1 \ldots )} } \cr
& & \frac{C_{(1 1 1 \ldots\mid 2 1 1 \ldots )\ (2 1 1 \ldots\mid 1 2 1 \ldots )}^{(2 1 1 \ldots\mid 2 1 1 \ldots )} }{C_{(1 1 1 \ldots\mid 2 1 1 \ldots )\ (2 1 1 \ldots\mid 2 1 1 \ldots )}^{(2 1 1 \ldots\mid 1 2 1 \ldots )} } \frac{C_{(2 1 1 \ldots\mid 1 1 1 \ldots )\ (2 1 1 \ldots\mid 2 1 1 \ldots )}^{(1 1 1 \ldots\mid 2 1 1 \ldots )} }{C_{(2 1 1 \ldots\mid 1 1 1 \ldots )\ (1 1 1 \ldots\mid 2 1 1 \ldots )}^{(2 1 1 \ldots\mid 2 1 1 \ldots )} }
\eea

which leads to : 
\beq
N_{(1 2 1 \ldots\mid 1 2 1 \ldots )}^2 = \left(  \frac{\pi}{\epsilon} \right)^{4(2r-1)} \left( 1 + \mathcal{O}(\epsilon) \right)
\eeq

This way one obtain the square of the vertex operator normalizations. One has to be careful when taking the square root, and make an analytic continuation of $\sqrt{N^2_{\Phi}}$ as a fucntion of $\epsilon$.

\subsection{Results}

Now we know all the $\wb{r}{}$ constants we need. We note that $\phi_{(1 1 1 \ldots\mid 1 2 1 \ldots )}$ is the only slightly relevant field of $\wb{r}{(p)}$, with the following algebra constant : 
\beq
D_{(1 1 1 \ldots\mid 1 2 1 \ldots ) \ (1 1 1 \ldots\mid 1 2 1 \ldots ) \ (1 1 1 \ldots\mid 1 2 1 \ldots )} = \frac{2(2r-1)}{\sqrt{r(2r+1)}} 
\eeq 

This implies that the $\wb{r}{(p)}$, being perturbed by the field $\phi_{(1 1 1 \ldots\mid 1 2 1 \ldots )}$, flows towards $\wb{r}{(p-1)}$. This confirms the observation, 
made with the $SU(2)$ cosets \cite{ref11,ref12} and, more generally, with the cosets 
for the simply laced algebras \cite{ref16}, that the perturbation of a coset theory
caused by an appropriate operator drives $p$ to $p-\Delta p$, 
$\Delta p$ being equal to the shift parameter of the coset.

Armed with the $\wb{r}{}$ constants we deduce the following results (at leading order in $\epsilon$) : 

\begin{itemize}

\item from (\ref{eqa},\ref{eqb}) we get the full decomposition (\ref{decompA})  of the field $A$ : 
\beq
a = b = \frac{1}{\sqrt{2}}, \quad c=0
\eeq

\item and then the $\z{2r+1}{p}$ constants we need :
\bea
\mathcal{D}_{AAA} & = &  \frac{(2r-1)}{\sqrt{r(2r+1)}} \\
\mathcal{D}_{SSA} & = &  \sqrt{\frac{2r+1}{r}}
\eea

\end{itemize}

\section{Renormalization group flows for $\z{N}{p}$}

\subsection{Beta functions}

We have obtained the values of  $\mathcal{D}_{AAA}$ and $\mathcal{D}_{SSA}$ at leading order in $\epsilon$. 
The renormalization group equations for the couplings $g$ and $h$ are then given by :
\bea
\beta_g = \frac{dg}{d\xi}& = & 2 (2r+1) \epsilon \  g-4\sqrt{\frac{2r+1}{r}}\ g\ h \\
\beta_h = \frac{dh}{d\xi}& = & 2 (2r-1) \epsilon \  h-2\frac{(2r-1)}{\sqrt{r(2r+1)}}\ h^{2}-2\sqrt{\frac{2r+1}{r}}\ g^{2}
\eea

These are up to (including) the first non-trivial order of the perturbations in $g$ and $h$.

These equations derive from a potential : 
\bea 
\beta_g & = &\partial_g V(g,h) \\
\beta_h & = &\partial_h V(g,h)
\eea

with : 
\beq
V(g,h) = (2r+1) \epsilon \ g^2  + (2r-1)\epsilon\  h^2 - 2 \sqrt{\frac{2r+1}{r}}\ g^2 \ h - \frac{2}{3} \frac{(2r-1)}{\sqrt{r(2r+1)}} \  h^3 
\label{potential}
\eeq

This potential plays a central role in the renormalization group flows. 
Let us consider the function $c(g,h)$ defined by $ c(g,h) = c_0 - \frac{V(g.h)}{24}$ : this is the c-function introduced by Zamolodchikov, which decreases along the renormalization group flows, and coincide with the central charge at any fixed point.

 At this point we can directly analyse the presence of IR fixed points for the renormalization group, and predict the corresponfing central charges.

The phase diagram of constants $g$ and $h$ contains (Fig. 1.):

\begin{itemize}
\item the UV fixed point $g_{0}^{\ast}=h^{\ast}_{0}=0 $, which obviously corresponds to the theory $\z{N}{p}$ 

\item the IR fixed point on the $h$ axis: \beq (g^{\ast}_1,h^{\ast}_1) = (0,\sqrt{r(2r+1)\epsilon}) \eeq

\item two additional IR fixed points for non-vanishing values of the two couplings: 
\beq
\begin{array}{lll}
(g^{\ast}_2,h^{\ast}_2) & = & (\frac{1}{2}\sqrt{r(2r-1)}\epsilon,\frac{1}{2}\sqrt{r(2r+1)}\epsilon), \\ 
(g^{\ast}_3,h^{\ast}_3) & = & (-\frac{1}{2}\sqrt{r(2r-1)}\epsilon,\frac{1}{2}\sqrt{r(2r+1)}\epsilon).
\end{array}
\eeq

\end{itemize}

To identify what conformal theory we have at these IR fixed points, we evaluate the central charge using the potential (\ref{potential}). We find that the value of the central charge at the point $(g^{*}_1,h^{*}_1) = (0,\sqrt{r(2r+1)\epsilon})$ agrees with that of the theory $\z{N}{p-2}$. This fixed point was to be expected.

On the other hand, the appearance of two extra fixed points, 
$(g^{\ast}_{2}$, $h^{\ast}_{2})$ 
and ($g^{\ast}_{3}$, $h^{\ast}_{3}$), is somewhat surprising. By the value
of the central charge, the two critical points 
correspond to the theory $\z{N}{p-1}$.

\vbox{
\begin{center}
\includegraphics[scale=0.7]{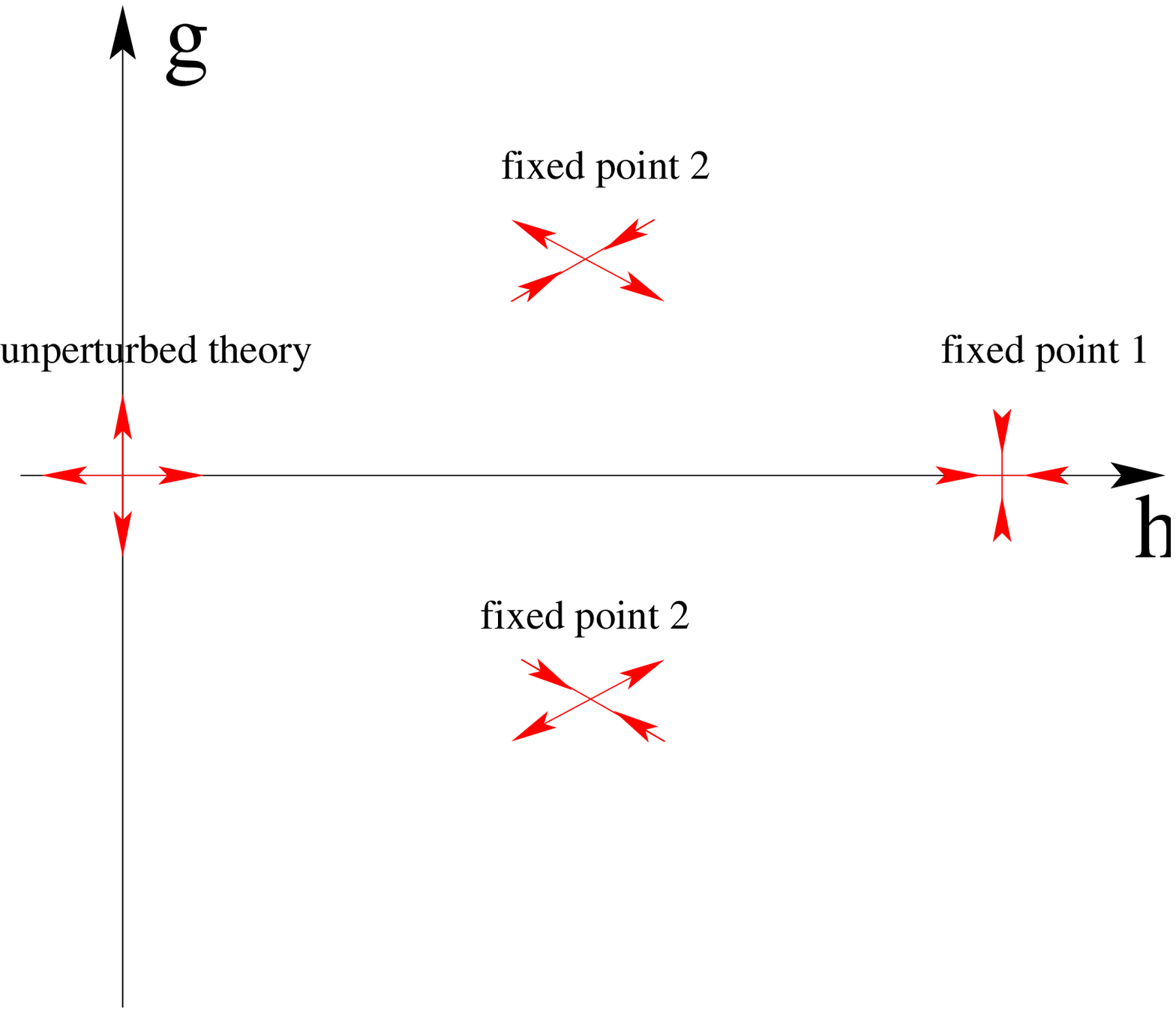}
\end{center}
\centerline{\em Fig. 1. Fixed points of the renormalization group.}
}

To check this statement, we evaluated the anomalous dimensions of some particular fields .

\subsection{Some gamma functions}

Our classification of these fixed points has further been
verified by calculating the critical dimensions at these points
of the operators $\Phi_{(1,n,\ldots \mid 1,n, \ldots)}$ and $\Phi_{(n,1,\ldots \mid n,1, \ldots)}$.

The gamma function giving the evolution of the dimension of a field of type $\Phi_{(\vec{n}\mid \vec{n})}$ is \cite{ref9,ref10} :
\beq
\frac{d2\Delta_{(\vec{n} \mid \vec{n})}}{d\xi} = \gamma_{(\vec{n} \mid \vec{n} )} = -4 h \mathcal{D}_{A(\vec{n} \mid \vec{n})(\vec{n} \mid \vec{n})} - 4 g \mathcal{D}_{S(\vec{n} \mid \vec{n})(\vec{n} \mid \vec{n})}
\eeq

We use the same techniques to evaluate the $\z{N}{p}$ constants $\mathcal{D}_{A(\vec{n} \mid \vec{n})(\vec{n} \mid \vec{n})}$ and  $\mathcal{D}_{S(\vec{n} \mid \vec{n})(\vec{n} \mid \vec{n})}$.  
First we identify the field $\Phi_{(\vec{n} \mid \vec{n})}$ in $\wb{r}{(p)} \otimes \wb{r}{(p+1)}$ : 
\beq
\Phi^{\z{N}{p}}_{(\vec{n} \mid \vec{n})} = \phi^{\wb{r}{(p)}}_{(\vec{n} \mid \vec{n})} \otimes \phi^{\wb{r}{(p+1)}}_{(\vec{n} \mid \vec{n})}
\label{ident}
\eeq

We note that (\ref{ident}) holds true only because $\Phi^{\z{N}{p}}_{(\vec{n} \mid \vec{n})}$ is always a neutral field.

We recall that 
\beq
 S = \mathcal{S}^{\z{N}{p}}_{(1 1 1 \ldots   \mid 3 1 1 \ldots  )} = \Phi^{\wb{r}{(p)}}_{(1 1 1 \ldots   \mid 2 1 1 \ldots  )} \otimes \Phi^{\wb{r}{(p+1)}}_{(2 1 1 \ldots   \mid 3 1 1 \ldots  )}
\label{decompS1}
\eeq

We can already see that $\mathcal{D}_{S(\vec{n} \mid \vec{n})(\vec{n} \mid \vec{n})} = 0 $ since it involves some  $\wb{r}{}$ constants with an odd number of fermionic screening operators :
\beq
\mathcal{D}_{S(\vec{n} \mid \vec{n})(\vec{n} \mid \vec{n})} = D_{(1 1 1 \ldots   \mid 2 1 1 \ldots  )(\vec{n} \mid \vec{n})(\vec{n} \mid \vec{n})} D_{(2 1 1 \ldots   \mid 3 1 1 \ldots  )(\vec{n} \mid \vec{n})(\vec{n} \mid \vec{n})} = 0
\eeq

Therefore $\gamma_{(\vec{n} \mid \vec{n} )}$ simplifies for singlets into :
\beq
\gamma_{(\vec{n} \mid \vec{n} )} = -4 h \mathcal{D}_{A(\vec{n} \mid \vec{n})(\vec{n} \mid \vec{n})}
\eeq

The problem now amounts to calculate $\mathcal{D}_{A(\vec{n} \mid \vec{n})(\vec{n} \mid \vec{n})}$. We use the expression : 
\bea
A = \Psi^{-1}_{-\frac{2}{N}} \mathcal{D}^{q=1}_{(1 1 1 \ldots \mid 1 2 1 \ldots)} & = & \frac{1}{\sqrt{2}} \  \Phi^{(\wb{r}{p})}_{(1 1 1 \ldots \mid 1 1 1 \ldots)}\otimes \Phi^{(\wb{r}{p+1})}_{(1 1 1 \ldots \mid 1 2 1 \ldots)} \cr
 & + &  \frac{1}{\sqrt{2}} \  \Phi^{(\wb{r}{p})}_{(1 1 1 \ldots \mid 1 2 1 \ldots)} \otimes \Phi^{(\wb{r}{p+1})}_{(1 2 1 \ldots \mid 1 2 1 \ldots)}
\eea

to obtain :
\beq
\mathcal{D}^{(p)}_{A(\vec{n} \mid \vec{n})(\vec{n} \mid \vec{n})} = 2 D^{(p+1)}_{(1 1 1 \ldots \mid 1 2 1 \ldots)(\vec{n} \mid \vec{n})(\vec{n} \mid \vec{n})}
\eeq

The integral corresponding to $D_{(1 1 1 \ldots \mid 1 2 1 \ldots)(\vec{n} \mid \vec{n})(\vec{n} \mid \vec{n})}$ has been estimated in the following cases : 

\begin{itemize}

\item $\vec{n} = (n 1 1 \ldots) $ :  $D_{(1 1 1 \ldots \mid 1 2 1 \ldots)(\vec{n} \mid \vec{n})(\vec{n} \mid \vec{n})} =     \frac{(n-1)(2r+n-2)}{\sqrt{r(2r+1)}} \epsilon^2$

\item $\vec{n} = (1 n 1 \ldots) $ :  $D_{(1 1 1 \ldots \mid 1 2 1 \ldots)(\vec{n} \mid \vec{n})(\vec{n} \mid \vec{n})} = \frac{2(n-1)(2r+n-3)}{\sqrt{r(2r+1)}} \epsilon^2$

\end{itemize}

So in these 2 cases the $\gamma$ function becomes : 

\begin{itemize}

\item 
$\gamma_{(n 1 1 \ldots \mid n 1 1 \ldots )} = -8 h \epsilon^2 \frac{(n-1)(2r+n-2)}{\sqrt{r(2r+1)}}$

\item
$\gamma_{(1 n 1 \ldots \mid 1 n 1 \ldots )} = -8 h \epsilon^2 \frac{2(n-1)(2r+n-3)}{\sqrt{r(2r+1)}}$

\end{itemize}

These values are in agreement with the statement that the field $\Phi^{(p)}_{(\vec{n} \mid \vec{n} )}$ flows towards $\Phi^{(p-k)}_{(\vec{n} \mid \vec{n} )}$: 
\beq
\Phi^{(p)}_{(\vec{n} \mid \vec{n} )} \rightarrow \Phi^{(p-k)}_{(\vec{n} \mid \vec{n} )} 
\eeq

with $k = \left\{ \begin{array}{l}  \textrm{2 at the fixed point 1} \\ \textrm{1 at fixed points 2 and 3} \end{array} \right.$

\section{Discussion}

In this paper, we have studied the effect of two slightly relevant perturbations for the second parafermionic theory $\z{N}{p}$, and we have found three fixed points. We have identified the corresponding conformal theories by evaluating the value of the central charge and the anomalous dimensions of some fields at these points. One of them is described by the expected $\z{N}{p-2}$ parafermionic theory. This confirms the observation, 
made with the $SU(2)$ cosets \cite{ref11,ref12} and, more generally, with the cosets 
for the simply laced algebras \cite{ref16}, that the perturbation of a coset theory
caused by an appropriate operator drives $p$ to $p-\Delta p$, 
$\Delta p$ being equal to the shift parameter of the coset.
In our case the shift parameter of the coset is equal to 2, eq.(\ref{coset_ZN}). 
Note that the algebra $B_{r}\equiv SO(2r+1)$ is not a simply laced one.

On the other hand, the appearance of two extra fixed points, corresponding to the theory $\z{N}{p-1}$, is somewhat surprising. 


\vbox{
\begin{center}
\includegraphics{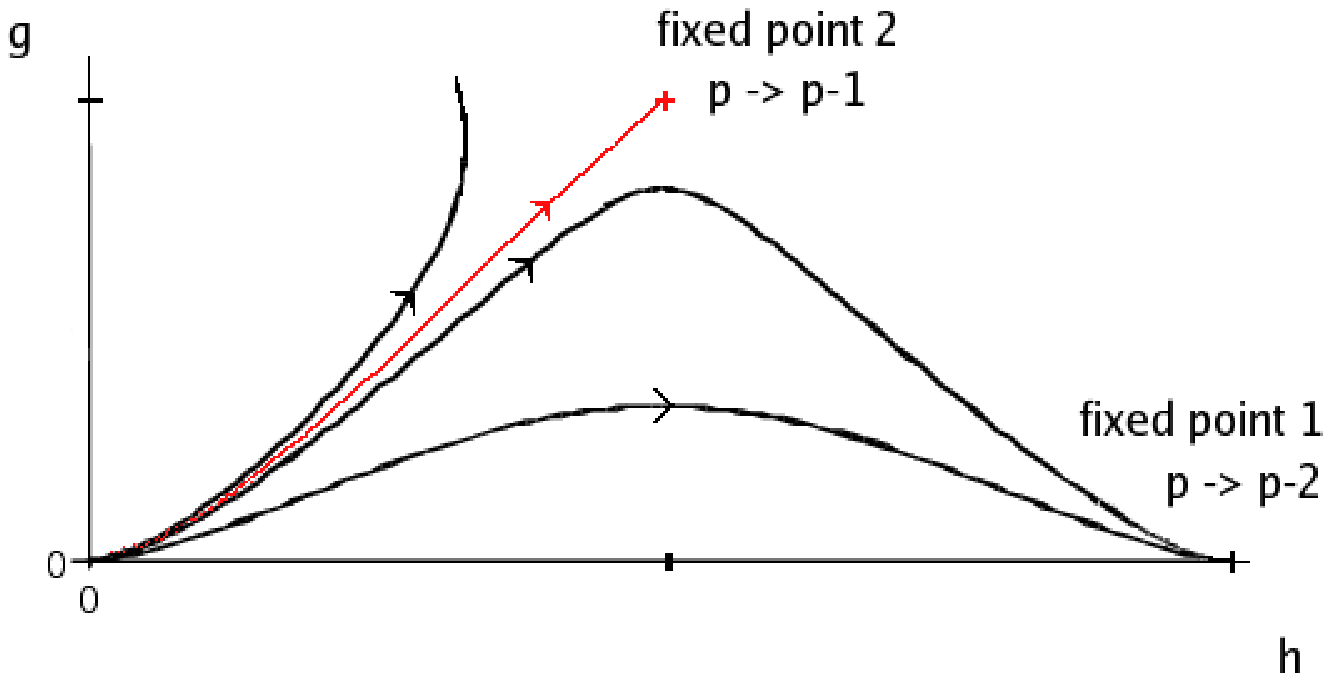}
\end{center}
\centerline{\em Fig. 2. Renormalization group flows. They are symmetrical with respect to $g \rightarrow - g$.}
}

We observe that such additional fixed points do not appear 
in the parafermionic model $Z_{3}^{(2)}$: the second $Z_{3}$ parafermionic theory 
with $\Delta_{\Psi}=4/3$ \cite{ref17}. This model could be realized 
with the $SU(2)$ cosets and its perturbations 
with two slightly relevant operators have been analysed in \cite{ref11,ref12}.

These two additionnal fixed points would have remained unseen if we had perturbated with the field $A$ alone. 

We can compare these results with those obtained for the second parafermionic theory $\z{N}{p}$ with $N$ even. Such theories are symmetric cosets on simply laced Lie algebras. The perturbation with one particular relevant field of the parafermionic theory $\z{N}{p}$ with $N$ even has already been treated in \cite{ref16}. The only fixed point obtained correspond to $\z{N}{p-2}$. On the basis of our present results, one could expect the presence of a second slightly relevant field and the existence of two additional fixed points corresponding to $\z{N}{p-1}$. This will be examined in \cite{ref18}.

\appendix

\section{A brief review of $\z{N}{p}$}
\label{appendixZN} 

The details of the second parafermionic theories $\z{N}{p}$ with $N$ odd can be found in \cite{ref6}.

The chiral algebra is made of $N-1$ parafermionic currents $\Psi^k$ (with $k=1,2,\dots,N-1$), and their operator product expansion is : 

\bea
\Psi^k \times \Psi^{k'} & \rightarrow & \Psi^{k+k'}\\
\Psi^k \times \Psi^{-k} & \rightarrow & \mathbb{I}
\eea

Note that in the above equation the $Z_N$-charges $k$ are defined modulo N. 

The currents have dimension :

\beq
\Delta_k = \Delta_{N-k} = \frac{2k(N-k)}{N}
\eeq

This implies the value of the central charge :

\beq
c=(N-1)\left(1-\frac{N(N-2)}{p(p+2)} \right)
\eeq

In general, the parafermionic algebra primaries of the second $Z_{N}$ conformal theory are labeled by 2 vectors ($\vec{n},\vec{n}'$) corresponfing to two ($\alpha_{+}$ and $\alpha_{-}$) lattices of the $B_{r}$ classical Lie algebra \cite{ref6}:
\beq
\Phi_{(\vec{n}\mid \vec{n}')} = \Phi_{(\underbrace{n_1,n_2, \dots}_{\alpha_{+}} \mid \underbrace{n_1',n_2',\dots}_{\alpha_{-}})}
\eeq
The first and second vector of indices correspond respectively to the  $\alpha_{+}$ and $\alpha_{-}$ $B_{r}$ lattices.
$\alpha_{+}$ and $\alpha_{-}$ are the usual Coulomb gas type parameters. 

The conformal dimension of primary operators takes the form : 

\beq
\Delta_{(\vec{n}\mid \vec{n}')} = \frac{ \left( \vec{n}(p+2) - \vec{n}'p \right)^2 - 4\vec{\rho}^2}{4p(p+2)} + B_Q
\label{KacZN}
\eeq

Where 

\beq
\begin{array}{lllll}
\vec{n} & = & (n_1,n_2,\ldots,n_r) & = &{\displaystyle \sum_{i=1}^{r}} n_i \vec{\omega}_i ,\\ 
\vec{n}' & = & (n'_1,n'_2,\ldots,n_r') & = & {\displaystyle \sum_{i=1}^{r}} n_i' \vec{\omega}_i 
\end{array}
\eeq

and $\vec{\omega}_i$, $i=1,\ldots r$ are the fundamental weights of the Lie algebra $B_r$

$B_Q$ in Eq.(\ref{KacZN}) is the "boundary term" which depends on the $Z_N$ charge $Q=2q$ mod $N$:

\beq
B_Q = \frac{Q(N-2Q)}{4N}, \quad Q = 0,1,2,\dots,\frac{N-1}{2}.
\eeq

The $Q$ charge of $Z_{N}$ takes values in $Z_{N}$, so that in the Kac table of this theory one finds the $Z_{N}^{(2)}$ neutral fields, of $Q=0$, the $Q = \pm 1,\pm 2,\dots,\pm \frac{N-1}{2} $ doublets, and the $Z_{2}$ disorder fields, with boudary term $B_R = \frac{1}{16}\left \lfloor \frac{N-1}{2} \right \rfloor$. 

Introducing $x_a = n_a - n_a'$ for $a=1,2,\dots r-1$ and $x_r = \frac{n_r-n_r'}{2}$, the doublet charge of the primary operator $\Phi_{(\vec{n}\mid \vec{n}')}$ is given by \cite{ref8} :

\beq
Q(\vec{n}-\vec{n}') = \sum_{a=1}^{r} \left[ \left(\sum_{b=a}^{r} x_b \right) \textrm{mod} 2     \right]
\eeq

We note that $\Phi_{(\vec{n}\mid \vec{n}')}$ is a disorder operator if $Q(\vec{n}-\vec{n}')$ is not an integer.

\section{Slightly relevant descendants of a doublet in the $\z{N}{p}$ theory}
\label{appendix_pertinentfields}

By fundamental descendant we mean a field that is still Virasoro primary. The doublets $\mathcal{D}^{Q=2q}$, $Q = 0,1,\ldots \frac{N-1}{2}$ have a non trivial boundary term in their dimension,
Any $Z_N$ fundamental descendant $A = \Psi^{q_1}_{-x_1} \ldots \Psi^{q_n}_{-x_n}D^{Q}$ that satisfies the neutrality condition $\sum_i 2q_i + Q = 0 $ mod $N$ necessarily has a gap $ \sum_i x_i $ equal to the fundamental gap $ \delta_Q = \frac{Q(Q-N)}{2N}$ mod[1]. The conformal dimension of such a descendant is : 

\beq
\Delta^{A}_{(\vec{n} \mid \vec{n}')}  =  \frac{((p+2)\vec{n} - p\vec{n}')^2 - 4\vec{\rho}^2}{4p(p+2)} + B_Q + \delta_Q
\eeq

where $\delta_{Q}$ is the fundamental gap $\delta_{Q} = \frac{Q(Q-N)}{2N}$ mod 1, and $B_Q$ is the boundary term $B_Q = \frac{Q(N-2Q)}{4N}$

Since we want $\Delta^{A}_{(\vec{n} \mid \vec{n}')}$ to be smaller than 1, $Q$ must obey $B_Q + \delta_Q < 1 $. In that case one can verify that 
$B_Q + \delta_Q = \frac{3Q}{4} $ mod 1.

We will denote the doublet as $D^{Q}_{(1,1,\ldots) \mid 1+n_1,1+n_2,  \ldots)}$. The dimension of $A_{(1,1,\ldots \mid 1+n_1,1+n_2,  \ldots)}$ then reads :

\beq
\Delta^{A}_{(1,1,\ldots \mid 1+n_1,1+n_2,  \ldots)}  =  \frac{\vec{n}^2}{4} + B_Q + \delta_Q +\mathcal{O}(\epsilon)
\eeq

Since $n_r$ is even for a doublet, we will redefine $n_r \rightarrow 2n_r$ for the sake of simplicity. We get : 

\beq
\vec{n}^2 =\left( \sum_{i=1}^{r} n_i \vec{\omega}_i \right)^2 = n_r^2 + (n_r + n_{r-1})^2 + (n_r + n_{r-1} +n_{r-2})^2 + \ldots + (n_r + \ldots +n_1)^2
\eeq

The condition $\Delta < 1$, which implies $\frac{\vec{n}^2}{4} < 1$, has the following solutions : 

\begin{itemize}

\item $\vec{n} = (1,0,0,0, \dots 0)$ which corresponds to $Q=1$

\item $\vec{n} = (0,1,0,0, \dots 0)$ which is a $Q=2$ doublet

\item $\vec{n} = (0,0,1,0, \dots 0)$ : a $Q=3$ doublet

\end{itemize}

This corresponds to the following admissible doublets :

\begin{itemize}

\item a $Q = 1$  $\left( q=\frac{N-1}{2} \right) $ doublet : $\mathcal{D}^{Q=1}_{(1,1,1,\ldots \mid 2,1,1,\ldots)}$

\item a $Q = 2$ $(q=1)$ doublet : $ \mathcal{D}^{Q=2}_{(1,1,1,\ldots \mid 1,2,1,\ldots)}$
\item and a $Q = 3$ $\left(q =\frac{N-3}{2}\right)$ doublet:  $\mathcal{D}^{Q=3}_{(1,1,1,\ldots \mid 1,1,2,\ldots)}$

\end{itemize}

These fields are the fundamental doublets of charges $Q=1,2,3$ \cite{ref6} : they correspond to the fields with the highest degenerate descendant. In that sense they have less descendants than the general doublet with the corresponding charge. The analysis of the degeneracy conditions leads to the following results : 

\begin{itemize}

\item $\mathcal{D}^{Q=1}_{(1,1,1,\ldots \mid 2,1,1,\ldots)}$ has no neutral descendant at level $\delta_1 = \frac{N+1}{2N}$

\item $ \mathcal{D}^{Q=2}_{(1,1,1,\ldots \mid 1,2,1,\ldots)}$ has one single neutral descendant at level $\delta_2 = \frac{2}{N}$. It is 
$A = \psi_{-\frac{2}{N}}^{-1}\mathcal{D}^{Q=2}_{(1,1,1,\ldots \mid 1,2,1,\ldots)} = \psi_{-\frac{2}{N}}^{1}\mathcal{D}^{Q=-2}_{(1,1,1,\ldots \mid 1,2,1,\ldots)} $

\item we conjecture, from the identification with $\wb{r}{(p)} \otimes \wb{r}{(p+1)}$, that $\mathcal{D}^{Q=3}_{(1,1,1,\ldots \mid 1,1,2,\ldots)}$
has no fundamental neutral descendant either. It has been explicitely checked in the case $N=7$. 
\end{itemize}

Finally there is one single neutral descendant of a doublet, with a trivial $\alpha_+$ side, that is slightly relevant : 

\beq 
A = \psi_{-\frac{2}{N}}^{-1}\mathcal{D}^{Q=2}_{(1,1,1,\ldots \mid 1,2,1,\ldots)}
\eeq

\section{Conventions for the $B_{r}$ Lie algebra}
\label{convbr}
In this paper we have adopted the following normalization conventions for the roots and weights. 

The simple roots are given by the Cartan matrix $A_{ij} = {\vec{e}_i}.\vec{e}_j \check{}  $, where $\vec{e}_j \check{} =  2 \vec{e}_j / \vec{e}_j^{\phantom{0}2}$ is the coroot of $\vec{e}_j$ : 

\beq
A = \left( \begin{array}{cccccc}
 2 & -1 &  0 & \ldots & 0 & 0\\
-1 &  2 & -1 & \ldots & 0 & 0\\
 0 & -1 &  2 & \ldots & 0 & 0 \\
 \vdots & \vdots   & \vdots & \ddots & \vdots & \vdots \\
0 & 0 & 0 & \ldots & 2 & -2\\
0 & 0 & 0 & \ldots & -1 & 2
\end{array} \right)
\eeq
$B_r$ is non simply laced since there are $r-1$ long roots : $\vec{e}_i^{\phantom{0}2} = 2 \textrm{ for } i=1\dots r-1$ and one short root $\vec{e}_r^{\phantom{0}2} = 1$.

The fundamental weights $\vec{\omega}_i$ form the base dual to the simple root basis : $\vec{\omega}_i . \vec{e}_j \check{} = \delta_{ij}$. 
The Cartan matrix is the transformation matrix relating the two basis $\{\vec{e}_{i} \}$ and $\{\vec{\omega}_{j} \}$ :

\beq
\vec{e}_i = A_{ij} \vec{\omega}_j
\eeq

The scalar product of weights can be expressed in terms of the symmetric quadratic form $\omega_{ij}$ : 

\beq
\omega_{ij} = \vec{\omega}_i . \vec{\omega}_j
\eeq

\beq
\omega = \frac{1}{2} \left( \begin{array}{cccccc}
 2 & 2 &  2 & \ldots & 2 & 1\\
 2 &  4 & 4 & \ldots & 4 & 2\\
 2 & 4 &  6 & \ldots & 6 & 3 \\
 \vdots & \vdots   & \vdots & \ddots & \vdots & \vdots \\
2 & 4 & 6 & \ldots & 2(r-1) & r-1\\
1 & 2 & 3 & \ldots & r-1 & r / 2
\end{array} \right)
\eeq
which correspond to :

\bea
\omega_{ij} = i, & i\leq j < r; \\
\omega_{in} = \frac{i}{2}, & i<r; \\ 
\omega_{nn} = \frac{n}{4} &
\eea

We will also introduce the Weyl vector : 
\beq
\vec{\rho} = \sum_i \vec{\omega}_i = (1,1,\dots,1)
\eeq

\section{The $\wb{r}{}$ Coulomb gas}
\label{appendixCG}

The $\wb{r}{}$ theories have been defined in \cite{ref14} through their Coulomb Gas. It is made of $r$ bosonic fields ${\varphi}_i$, quantized with 
a background charge and the Ising model fields: $\Psi$ (free fermion) and $\sigma$ (spin operator).

The screening operators are given by : 

\bea
V^{(a)}_{\pm}(z,\bar{z}) & = & \  : \mathrm{exp}\left(i\alpha_{\pm}\frac{\vec{e}_{a}}{\sqrt{2}}.\vec{\varphi}(z,\bar{z})\right) : \quad , \quad a =1 \dots r-1 \\
V^{(r)}_{\pm}(z,\bar{z}) & = & \  : \psi(z)\overline{\psi}(\bar{z})\mathrm{exp}\left(i\alpha_{\pm}\frac{\vec{e}_{a}}{\sqrt{2}}.\vec{\varphi}(z,\bar{z})\right) : \\
\alpha_{+} & = & \sqrt{\frac{p+1}{p}}\\
\alpha_{-} & = & -\sqrt{\frac{p}{p+1}}
\eea
The normalization adopted for the bosonic fields is  $\left< \varphi_i(z,\bar{z}) \varphi_j(z',\bar{z}')  \right> = - 2 \delta_{i,j} \mathrm{log}(|z-z'|^2)$.
$\vec{e}_a$ are the simple roots of $B_r$ (cf Appendix \ref{convbr}).

The vertex operators representing primary fields : 

\bea
V_{\vec{\beta}_{(\vec{n} \mid \vec{n}')}}(z,\bar{z})  = & : \mathrm{exp}\left(i \vec{\beta}_{(\vec{n} \mid \vec{n}')}.\vec{\varphi}(z,\bar{z})\right) : & \textrm{if $n_r-n_r'$ is even}\\
V_{\vec{\beta}_{(\vec{n} \mid \vec{n}')}}(z,\bar{z})  = & \sigma(z,\bar{z}) \ : \mathrm{exp}\left(i \vec{\beta}_{(\vec{n} \mid \vec{n}')}.\vec{\varphi}(z,\bar{z})\right) : & \textrm{if $n_r-n_r'$ is odd} 
\label{vertex}
\eea

with weight $\vec{\beta}_{(\vec{n} \mid \vec{n}')}  =  \sum_{i=1}^{r} \left(\frac{1-n_i}{2}\alpha_{+}  + \frac{1-n_i'}{2}\alpha_{-}  \right) \sqrt{2}  \vec{\omega}_i$. 

$\vec{\omega}_i$ are the fundamental weights of $B_r$. (cf Appendix \ref{convbr}).

The background charge is : 

\beq
\vec{\alpha}_{0} = \frac{\alpha_{+} + \alpha_{-}}{2} \sqrt{2} \vec{\rho} 
\eeq

with $\vec{\rho} = \sum_i \vec{\omega}_i$ is the Weyl vector. 

We can check the value of the central charge : 

\beq
c = c_{\mathrm{bosons}} + c_{\mathrm{ising}} = \left(r - 24 \vec{\alpha}_{0}^2 \right) + 1/2 = \left( r+ \frac{1}{2} \right) \left(1- \frac{2r(2r-1)}{p(p+1)} \right)
\eeq

obtained with the stress-energy tensor :

\beq
T(z) = -\frac{1}{4} : \partial_z \vec{\varphi}(z) . \partial_z \vec{\varphi}(z) : +i \vec{\alpha}_{0}.\partial_z^2\vec{\varphi}(z) + \frac{1}{2}: \partial \psi \psi : 
\eeq

The dimension of the $V_{\vec{\beta}}(z,\bar{z})$ is 

\beq
\Delta_{\vec{\beta}} = \vec{\beta}^2 - 2\vec{\beta}.\vec{\alpha}_{0} = \left(\vec{\beta} - \vec{\alpha}_{0}\right)^2 - \vec{\alpha}_{0}^2 = \Delta_{2\vec{\alpha_0} - \vec{\beta}}
\label{CGdimbeta}
\eeq

so that a primary field has two representations : $V_{\vec{\beta}}$ and $\overline{V}_{\vec{\beta}} = V_{2\vec{\alpha_0} - \vec{\beta}} $.

Using (\ref{vertex}) and (\ref{CGdimbeta}), we find the Kac formula of $\wb{r}{(p)}$ : 

\bea
\Delta^{(p)}_{(\vec{n} \mid \vec{n}')} = & \frac{\left( (p+1)\vec{n}-p\vec{n}'\right)^2 - \vec{\rho}^2}{2p(p+1)} & \textrm{if $n_r-n_r'$ is even (Neveu-Schwarz field)}\\
\Delta^{(p)}_{(\vec{n} \mid \vec{n}')} = & \frac{\left( (p+1)\vec{n}-p\vec{n}'\right)^2 - \vec{\rho}^2}{2p(p+1)} + \frac{1}{16} & \textrm{if $n_r-n_r'$ is odd (Ramond field)}
\eea


\begin{thebibliography}{99}

\bibitem{ref1}       V.~A.~Fateev and A.~B.~Zamolodchikov,
                     Sov.~Phys.~JETP {\bf 62} (1985) 215.

\bibitem{ref2}       N.~Read and E.~Rezayi, 
                     Phys.~Rev. B {\bf 59} (1999) 8084.

\bibitem{ref3}       H.~Saleur,
                     Comm.~Math.~Phys. {\bf 132} (1990) 657; Nucl.~Phys. B {\bf 360} (1991) 219.

\bibitem{ref4}       D.~Gepner,
                     Nucl.~Phys. B {\bf 296} (1988) 757. 


\bibitem{ref5}       Vl.~S.~Dotsenko, J.~L.~Jacobsen and R.~Santachiara,
                     Nucl.~Phys. B {\bf 656} (2003) 259.
                     
\bibitem{ref6}       Vl.~S.~Dotsenko, J.~L.~Jacobsen and R.~Santachiara, 
                     Nucl.~Phys. B {\bf 664} (2003) 477. 
                     
\bibitem{ref7}       Vl.~S.~Dotsenko, J.~L.~Jacobsen and R.~Santachiara, 
                     Phys.~Lett. B {\bf 584} (2004) 186. 
                     
\bibitem{ref8}       Vl.~S.~Dotsenko, J.~L.~Jacobsen and R.~Santachiara, 
                     Nucl.~Phys. B {\bf 679} (2004) 464.

\bibitem{refletter}  Vl.~S.~Dotsenko, B.~Estienne,
		     Phys.~Lett. B {\bf 643} (2006) 362.

\bibitem{ref9}       A.~B.~Zamolodchikov,
                     Sov.~Phys.~JETP Lett. {\bf 43} (1986) 730;
                     Sov.~J.~Nucl.~Phys. {\bf 46} (1987) 1090.

\bibitem{ref10}      A.~W.~Ludwig and J.~L.~Cardy, 
                       Nucl.~Phys. B {\bf 285} (1987) 687.
                        
\bibitem{ref11}      C.~Crnkovi\'c, G.~M.~Sotkov, and M.~Stanishkov, 
                             Phys.Lett.B226:297,1989

\bibitem{ref12}      C.~Crnkovi\'c, R.~Paunov, G.~M.~Sotkov, and M.~Stanishkov,
                             Nucl.Phys.B336:637,1990

\bibitem{ref13}      P.~Goddard and A.~Schwimmer,
                       Phys.~Lett. B {\bf 206} (1988) 62.

\bibitem{ref14}       V.~A.~Fateev and S.~I.~Luk'yanov,
                      Sov.Sci.Rev.A Phys.Vol{\bf 15} (1990) 1.


\bibitem{ref16}      V.~A.~Fateev, 
                       Phys.~Lett. B {\bf 324} (1994) 45

\bibitem{ref17}       V.~A.~Fateev and A.~B.~Zamolodchikov,
                     Theot.~Math.~Phys.{\bf 71} (1987) 451

\bibitem{ref18}      B.~Estienne,
                      in preparation

\bibitem{ref19}        Vl.~S.~Dotsenko and V.~A.~Fateev,
                     Nucl.Phys.B {\bf 251} 691:734 (1984)

\end{thebibliography}
\end{document}